\documentclass{MITcsail}

\definecolor{CiteGray}{HTML}{777777} 
\definecolor{CiteSlate}{HTML}{475569}  
\hypersetup{
  linkcolor=CiteGray,            
  citecolor=CiteGray,         
  urlcolor=CiteGray,          
  linktoc=page,                
}

\title{Picturing Perceptions: An Open-Source Toolkit to Uncover Bias in Humans and Machines}

\author
{Saurabh Khanna~$^{1, 2}$\footnote{Correspondence E-mail: s.khanna@uva.nl}, Zhijun Chen~$^{1}$, Chei Billedo~$^{1}$, Jiayi Yan~$^{1}$, Irene van Driel~$^{1}$, Alex Barco Martelo~$^{3}$, Hugo Moreda Cartagena~$^{3}$, Haizea González Atorrasagasti~$^{3}$, Markel Adánez Pérez~$^{3}$, Daniela An~$^{1}$, Qianyi Wang~$^{1}$, Xinkangrui Gao~$^{1,4}$, Lauren Taylor$^{1}$, Olga Eisele$^{1}$, Sindy Sumter$^{1}$\\
\vspace{1em} 
\normalfont{\small $^{1}$Amsterdam School of Communication Research, University of Amsterdam}\\
\normalfont{\small $^{2}$Pembroke College, University of Oxford}\\
\normalfont{\small $^{3}$Department of Computing, Electronics and Communication Technologies, University of Deusto}\\
\normalfont{\small $^{4}$Department of Media and Communication, City University of Hong Kong}\vspace{2em}
}

\begin{document}

\maketitle
\thispagestyle{firstpagestyle} 

\vspace{-1cm}

\onehalfspacing

\begin{abstract}
Bias in human judgment and artificial intelligence systems poses critical challenges across consequential domains like hiring, loans, and criminal justice. However, traditional bias measurement tools face fundamental limitations: they struggle to capture intersectional identities, cannot evaluate AI systems, lack grounding in demographic reality, and remain vulnerable to social desirability effects. We introduce \textit{PictoPercept}, an open-source\footnote{PictoPercept is being developed as a fully open-source toolkit on \href{https://github.com/invisibleinfo/pictopercept/}{GitHub}.} toolkit that measures bias through visual forced-choice comparisons grounded in population-level benchmarks. Participants view pairs of normed facial photographs and assess who is more likely to have higher earnings, with selections compared against actual U.S. Bureau of Labor Statistics data. We validate \textit{PictoPercept} with a nationally representative sample of 283 American adults and assess GPT-5, a mainstream generative model, using identical stimuli. Our study reveals three key findings: First, participants dramatically underestimate Asian American earnings despite this group having the highest actual earnings, while overestimating Latino male and White male earnings. Second, ingroup favoritism is not universal as White males show clear ingroup bias, but Asian participants actually underestimate their own group's earnings. Third, GPT-5 exhibits substantially stronger biases than humans, with stark systematic underestimation of all female groups. These findings suggest that \textit{PictoPercept} enables unified bias assessment across human and AI systems while revealing systematic misperceptions that diverge from demographic reality.

\end{abstract}

\doublespacing

\section{Introduction}

Bias in human judgment and artificial intelligence systems has emerged as one of the most pressing challenges facing communication scientists \citep{mehrabi2021survey, friedler2019comparative}. Hiring algorithms that perpetuate gender disparities \citep{chen2023review} and criminal justice \citep{arowosegbe2023data} models that exhibit racial bias exemplify how the repercussions of undetected bias can be profound and widespread. Despite widespread acknowledgment of this issue, bias measurement continues to be a fundamental bottleneck. Traditional methodologies for assessing bias, applicable to both humans and machines, encounter significant methodological constraints that hinder both theoretical comprehension and practical intervention \citep{gawronski2006associative}.

Traditional survey-based measures of bias in interpersonal perception studies have long suffered from several critical weaknesses. They require separate instruments for each dimension of interest (such as gender, ethnicity, or age), making comprehensive assessment inefficient and expensive in terms of survey time. Question ordering can in itself systematically influence responses, introducing measurement artifacts difficult to disentangle from true attitudes \citep{schuman1996questions}. Self-report measures are particularly vulnerable to social desirability bias, as respondents may consciously or unconsciously adjust their answers to align with perceived norms \citep{crowne1960new}. Even more importantly, traditional surveys are fundamentally text-based and explicit, limiting their ability to capture the automatic, implicit processes that often drive biased judgments in real-world contexts. For artificial intelligence systems, which operate through fundamentally different and mechanical decision pathways, survey-based approaches are entirely inapplicable. 

The Implicit Association Test (IAT) was developed to address these limitations in the assessment of human bias by measuring automatic associations through response latencies rather than explicit self-reports \citep{greenwald1998measuring}. However, the IAT's factorial design -- comparing category combinations in separate blocks -- severely constrains its ability to capture intersectional bias across multiple dimensions simultaneously \citep{ghavami2013intersectional}. A measure that can, for instance, assess whether bias against Black women exceeds the additive effect of anti-Black and anti-female bias cannot be readily constructed within the IAT framework. Furthermore, the IAT's reliance on reaction time differences makes it unsuitable for assessing AI models, which either lack reaction times entirely or produce them through entirely different mechanisms than human cognitive processing. 

The lack of robust and standardised tools to quantify bias in artificial intelligence (AI) systems constitutes a persistent and critical methodological gap. Although algorithms are often portrayed as neutral and objective, their output can exhibit systematic and reproducible biases \citep{tolan2019fair}. AI models are typically trained on large-scale datasets that are shaped by human choices and practices, within which social biases are embedded and propagated \citep{howard2018ugly}. Consequently, these systems can mirror and amplify human biases, leading to algorithmic bias characterised by systematic, replicable errors that reinforce existing patterns of inequality. Therefore, a rigorous and continuous assessment of bias in AI-based decision-making processes is indispensable.

Beyond these measure-specific limitations in human assessments and algorithmic bias lies a more fundamental problem: current bias assessment tools operate in isolation from the ground realities. Neither surveys nor IATs systematically compare measured perceptions against actual population-level statistics from census data or labor force surveys \citep{blasi2006system}. This disconnect makes it difficult to distinguish between accurate perception of existing inequalities and biased distortion that either exaggerates or minimizes real-world disparities. Additionally, the gulf between human bias research (dominated by methods from cognitive and social psychology like surveys and IATs) and AI fairness research (focused on algorithmic audits and dataset analyses) has impeded cross-domain synthesis. Different communities study similar phenomena using incompatible methodologies, making direct comparisons almost impossible. Finally, proprietary measures and implementations of some of these tools limit reproducibility and hinder the cumulative development of knowledge that characterizes mature scientific fields \citep{open2015estimating}.

Given these challenges, we introduce \textit{PictoPercept}, a novel open-source methodological toolkit for measuring bias in both humans and artificial intelligence systems. Our approach leverages the richness and implicitness of visual stimuli rather than text-based queries, presenting participants (human or AI) with repeated forced-choice comparisons between pairs of normed facial photographs \citep{ma2015chicago}. By holding the question prompt constant (say, `Which person is more likely to have higher personal earnings?') while systematically varying the demographic characteristics of the faces presented, we capture bias signals across multiple intersecting dimensions simultaneously -- gender, ethnicity, and their interactions -- without the order effects that plague sequential questioning. Because both humans and AI models can respond to the same visual forced-choice task, \textit{PictoPercept} enables direct comparison of bias patterns across these two domains using a unified measurement framework.

This framework enables the assessment of bias relative to multiple classes of benchmarks, including idealised benchmarks (e.g., equal distributions across groups) and empirically grounded real-world benchmarks (e.g., population-level statistics). In the present study, we adopt the latter approach, employing data from the U.S. Bureau of Labor Statistics as an empirical reference standard \citep{us2021census}. Within this framework, we conceptualise bias as a systematic distortion in individuals’ perceptions of economic outcomes relative to these observed benchmarks. More precisely, bias is taken to occur when judgments about group-level earnings are systematically shaped by demographic cues and consequently diverge from empirically documented population patterns, producing consistent overestimation or underestimation. Our focus is thus not on bias as an overt prejudice or enacted discrimination, but rather on systematic distortions in social inference.

Our study has two primary objectives. First, we validate \textit{PictoPercept} as a psychometric instrument by examining its internal consistency, convergent validity with established bias measures, and divergent validity with unrelated constructs such as personality traits and social desirability \citep{campbell1959convergent}. This validation is essential to establish that our novel method measures bias specifically rather than general response tendencies. Second, we apply the validated tool to investigate earnings-related bias across eight self-reported demographic categories (Male/Female $\times$ Asian/Black/Latino/White) in both human participants and the state-of-the-art GPT-5 AI language model. By comparing both human and AI judgments against actual US labor force statistics \citep{bls2025earnings}, we assess the magnitude and direction of perceptual distortions, test for intersectional effects that exceed additive predictions, and directly compare bias patterns between biological and artificial intelligence. 

We find that PictoPercept reveals systematic earnings perception biases across all demographic groups, with three key findings. First, participants dramatically underestimated Asian American earnings despite this group having the highest actual earnings, while overestimating Latino male and White male earnings. Second, ingroup favoritism was not universal -- White males showed clear ingroup bias, but Asian participants actually underestimated their own group's earnings. Third, GPT-5 exhibited substantially stronger biases than humans, with extreme systematic underestimation of all female groups and overestimation of male groups. These patterns held across diverse subgroups, though with varying magnitudes based on participant characteristics.

Our study makes a theoretical contribution by building on the methodological advancements outlined above to enhance existing approaches to bias measurement through a more efficient, integrated, multidimensional, and empirically grounded framework. This advancement strengthens the theoretical understanding of how implicit and intersectional biases can be systematically identified and analyzed beyond the constraints of traditional instruments. The study also makes a societal contribution by introducing an open-source and transparent toolkit that facilitates reliable detection, comparison, and mitigation of bias across both human and AI judgments. In doing so, it supports efforts to promote fairness, accountability, and inclusivity in social and computational decision systems.


\section{Theoretical Framework and Prior Research}

\subsection{Prior Research on Bias Measurement}

\subsubsection{The Nature and Challenge of Measuring Bias}

Bias can be defined as a systematic deviation from rational judgment that arises when people rely on mental shortcuts, or heuristics, to make decisions under uncertainty  \citep{tversky1974judgment,korteling2018neural}. The human mind does not passively record information but actively constructs it through learned associations and categorical shortcuts \citep{gazzaniga1998mind,bar2007proactive}. Rather than representing mere opinion or belief, bias is a fundamental perceptual process through which the mind organizes and interprets information
\citep{korteling2018neural}. Through repeated exposure and learning, these associative patterns become ingrained cognitive structures that guide perception across diverse contexts -- whether evaluating people, objects, or events. Understanding bias as a perceptual phenomenon, rather than simply an attitudinal one, has profound implications for how it should be measured.

Human judgments rarely operate on single dimensions in isolation. Configural Theory \citep{asch1946forming} and Information Integration Theory
\citep{anderson1981foundations}
propose that people evaluate stimuli holistically, combining multiple pieces of information into unified impressions rather than assessing each attribute independently. These theoretical frameworks suggest that bias is inherently multidimensional, emerging from the interplay among attributes rather than from isolated judgments about any single characteristic. Intersectionality theory extends this multidimensional perspective to social identities, proposing that characteristics such as gender, ethnicity, and age combine to produce unique patterns of stereotype and discrimination that cannot be understood by examining each dimension in isolation \citep{crenshaw2013demarginalizing}. In everyday social perception, people do not judge `gender' or `ethnicity' as separate attributes but rather encounter them as integrated wholes -- `Asian woman,' `Black man,' `White woman' -- each carrying distinct social meanings and associations \citep{ghavami2013intersectional}. Traditional measures of bias, however, typically treat these dimensions independently or at most in simple additive models, thereby missing the complex interactive effects that characterize real-world social cognition. Recognizing bias as an interactive, multidimensional phenomenon underscores the fundamental need for measurement tools capable of capturing overlapping patterns across multiple dimensions simultaneously.


Bias in artificial intelligence systems emerges when algorithms learn from data that encode existing social and structural inequalities
\citep{ntoutsi2020bias,zhang2018mitigating}
. Machine learning models detect and replicate statistical patterns within their training data; if those data include biased or unbalanced representations of demographic groups, the resulting models reproduce and sometimes amplify these distortions 
\citep{zhang2018mitigating}
. Such bias can also stem from subjective design choices in data labeling, model architecture, or evaluation criteria, embedding human value judgments into computational systems 
\citep{vangiffen2022overcoming}
. However, measuring bias in AI is particularly challenging due to the complexity and opacity of models. Deep learning systems often operate as "black boxes," where countless internal parameters interact in non-transparent ways, making it difficult to identify which features, data patterns, or decision layers produce biased outcomes 
\citep{tjoa2020survey,linardatos2020explainable,li2022interpretable}
. And because AI systems generate outputs almost instantaneously rather than through timed cognitive processes, reaction-time–based measures such as the Implicit Association Test (IAT) cannot be meaningfully applied to assess their bias.

\subsubsection{Evolution of Bias Measurement Approaches}

Early attempts to measure bias primarily relied on self-report instruments such as attitude surveys and Likert-scale ratings, where participants explicitly evaluated social groups, traits, or preferences \citep{axt2018best,kreitchmann2019controlling,hofmann2005meta}. While these explicit measures offered insight into conscious beliefs and openly endorsed attitudes, they suffered from significant limitations. Most prominently, self-reports are vulnerable to social desirability bias and self-presentation effects -- participants often consciously or unconsciously modify their responses to align with perceived social norms rather than revealing their genuine evaluations \citep{crowne1960new,caputo2017social,kreitchmann2019controlling,kwak2019measuring}. Additionally, explicit measures may fail to capture attitudes that operate outside conscious awareness or that individuals genuinely believe they do not hold.

To overcome these limitations and access automatic associations beyond conscious control, researchers developed implicit measurement techniques. The Implicit Association Test (IAT) emerged as the dominant approach for measuring implicit bias, assessing automatic associations through differences in categorization speed when individuals pair concepts under different conditions \citep{greenwald1998measuring}. The IAT transformed bias research by demonstrating that even individuals who consciously endorse egalitarian values can display measurable implicit preferences that diverge from their explicit self-reports. Meta-analytic evidence indicates that IAT scores predict discrimination and biased behavior beyond explicit measures, particularly in contexts where social desirability concerns are high or where behaviour is spontaneous rather than deliberative \citep{greenwald2009understanding}. Despite its theoretical importance and widespread adoption, the IAT faces several conceptual and methodological limitations. First, its binary categorical structure simplifies complex attitudes into two opposing categories, making it fundamentally unsuited to capture interactions among multiple dimensions such as gender and ethnicity simultaneously. The IAT's factorial design requires separate testing blocks for different attribute pairings, preventing the simultaneous assessment of intersectional combinations. Second, the IAT relies on text-based and symbolic stimuli (words and category labels) rather than the rich perceptual information through which bias naturally manifests in face-to-face social interaction. Third, its dependence on reaction-time differences restricts application to human participants who produce temporal response patterns; this measurement approach cannot be extended to artificial intelligence systems, which either lack reaction times entirely or produce them through fundamentally different computational mechanisms.

These limitations collectively reveal a critical gap in the bias measurement literature. Existing tools excel at demonstrating that bias exists and can distinguish between explicit and implicit forms of evaluation, but they struggle to capture several key features of bias as it operates in naturalistic social contexts. First, current measures typically cannot assess multiple intersecting dimensions of identity simultaneously, missing the nuanced patterns that characterize real-world discrimination. Second, they rely on abstract, text-based stimuli rather than the perceptual cues -- particularly visual facial information -- through which social categorization most commonly occurs. Third, existing measures produce relative indices of bias (such as preference for Group A over Group B) but lack grounding in objective benchmarks, making it difficult to distinguish between accurate perception of existing social inequalities and perceptual distortion that either exaggerates or minimizes real-world disparities \citep{blasi2006system}. Perhaps most critically, the methodological divide between human bias research (dominated by psychology methods like surveys and IATs) and AI fairness research (focused on algorithmic audits and dataset analyses) has impeded cross-domain synthesis \citep{friedler2019comparative}. Different research communities study similar phenomena using incompatible methodologies, making direct comparison between human and machine bias difficult. As artificial intelligence systems increasingly make consequential decisions in domains ranging from hiring to criminal justice, the inability to assess AI bias using the same measurement framework applied to human bias represents a significant barrier to understanding how human biases may be encoded in, amplified by, or potentially corrected through machine learning systems. Finally, proprietary measurement tools and closed-source implementations limit reproducibility and hinder the cumulative development of knowledge that characterizes mature scientific disciplines \citep{open2015estimating}.

\subsection{The PictoPercept Theoretical Framework}

Given the limitations of existing approaches, we propose a novel measurement framework -- \textit{PictoPercept} -- grounded in five core theoretical principles that address the identified gaps while remaining applicable to both human and artificial intelligence systems.

\subsubsection{Image-Based Measurement}

Bias tends to originate in perception and manifests primarily through visual social categorization in everyday contexts, and hence could be measured through perceptual experiences that resemble real-world judgment processes. When the conditions of a study reflect the real environments in which people think and behave, the design becomes more representative and supports more valid, generalizable conclusions \citep{brunswik2023perception,araujo2007ecological}. Traditional measurement tools -- whether text-based surveys or reaction-time tasks using word stimuli -- create artificial testing conditions that isolate bias from the rich perceptual contexts in which it naturally occurs. These abstract formats require participants to translate visual social experiences into linguistic categories and then respond to decontextualized verbal prompts, introducing layers of cognitive processing that may obscure or distort the immediate perceptual foundations of bias. In contrast, image-based measurement can offer superior validity by capturing bias within the same visual and associative processes that guide everyday social evaluation. Research on visual social cognition demonstrates that people form trait impressions -- including judgments of trustworthiness, competence, dominance, and attractiveness -- within milliseconds of exposure to facial photographs, through largely automatic and unconscious processing \citep{willis2006first}. These rapid face-based judgments influence consequential real-world decisions including voting behaviour, hiring choices, and legal outcomes \citep{todorov2005inferences}. The speed and automaticity of these processes suggest that they reflect evolutionarily perceptual mechanisms that prioritize rapid social categorization over deliberative reasoning \citep{kahneman2011thinking}.

Following this theoretical logic, the PictoPercept approach grounds bias measurement in the same rapid, intuitive perceptual processes that characterize spontaneous social evaluation. By presenting participants with photographic images of faces rather than category labels or verbal descriptions, the task elicits immediate visual impressions rather than reflective reasoning about abstract social groups. This design captures the phenomenology of bias as it manifests in everyday perception -- automatic, holistic, and tied to specific perceptual features rather than conscious categorical thinking. Thus, image-based bias measures not only align theoretically with the perceptual foundations of bias but also achieve higher ecological realism by capturing responses in a format that mirrors actual social cognition, thereby bridging the gap between laboratory assessment and real-world cognitive processes.

\subsubsection{Simultaneous Multidimensional Assessment}

A distinctive advantage of image-based measurement is its capacity to represent multiple social category cues simultaneously within a single stimulus, enabling bias assessment across intersecting identities rather than along isolated dimensions. Each photograph in the PictoPercept task inherently displays multiple categorical attributes -- gender expression, apparent ethnicity, approximate age, and various other facial characteristics -- without requiring the artificial decomposition of social identity into separate testing blocks. This simultaneous presentation reflects how social categories are actually encountered in everyday perception: people do not perceive `gender' and `ethnicity' as separate, sequential judgments but rather as integrated aspects of a unified social perception \citep{FREEMAN2020237}. Theoretically, this measurement structure aligns with intersectionality theory, which argues that overlapping social identities create unique patterns of privilege, discrimination, and social experience that cannot be understood through additive models \citep{crenshaw2013demarginalizing}. Social psychological research on intersectional stereotypes confirms that perceivers hold distinct mental representations for demographic combinations (e.g., `Asian woman,' `Latino man') that incorporate unique traits not derived from either single category alone \citep{ghavami2013intersectional}.

Methodologically, embedding intersectional information directly into visual stimuli allows PictoPercept to calculate separate bias indices for each demographic combination without requiring participants to complete separate testing blocks for each dimension or combination. Because every face naturally displays both gender and ethnicity simultaneously, patterns of image selection across trials reveal preferences for specific intersectional combinations. Moreover, this design enables detection of multiplicative or super-additive intersectional effects -- cases where bias against a particular combination exceeds the additive combination of biases against each constituent category. For instance, if participants show +10\% bias favouring White individuals and +15\% bias favouring men, intersectionality theory would suggest that White men might receive more than +25\% preference if intersectional amplification occurs. By capturing these nuanced, multidimensional patterns of bias within an integrated measurement framework, PictoPercept enables detection of intersectional dynamics that remain inaccessible to traditional single-dimension approaches.

\subsubsection{Reducing Social Desirability}

Bias assessment is inherently complicated by social desirability concerns and self-presentation motivations. When directly asked about their attitudes toward social groups, respondents often adjust their answers to appear unprejudiced and socially appropriate, regardless of their actual perceptual tendencies. Traditional Likert-scale measures are particularly vulnerable to this problem because the socially desirable response direction is typically transparent, enabling deliberate impression management \citep{fisher1993social}. To reduce social desirability effects while maintaining measurement validity, PictoPercept employs a comparative judgment format in which participants choose between two simultaneously presented stimuli rather than making absolute ratings or explicit category-based evaluations. When selecting rapidly between two facial photographs (`Which person is more likely to have higher personal earnings?'), participants reveal spontaneous perceptual impressions through their pattern of choices across repeated trials. Crucially, this task does not ask participants to explicitly endorse beliefs about social groups but rather to make repeated person-level comparisons that may be influenced by gender, ethnicity, or other relevant characteristics without requiring conscious acknowledgment of those influences.

Research on comparative judgment demonstrates that relative evaluations -- choosing between options rather than rating each independently -- capture psychological differences with greater reliability and precision than absolute rating scales \citep{goffin2011relative}. Comparative formats force participants to discriminate between stimuli on relevant dimensions, reducing response biases such as scale-use differences, acquiescence, and reference-point effects that plague rating scales. Participants may select faces based on perceptual cues without consciously categorizing those selections as reflecting gender or ethnic bias, allowing bias to emerge more naturally through patterns of comparative preference aggregated across many trials. This approach aligns with the principle that implicit bias is best captured through indirect measurement methods that minimize conscious deliberation and strategic responding \citep{sanbonmatsu1990role}. By embedding bias assessment within an ostensibly neutral comparative task and aggregating across numerous trials, PictoPercept captures systematic patterns in perceptual judgment while reducing participants' ability and motivation to consciously regulate their responses according to egalitarian self-presentation goals.

\subsubsection{Capturing System 1 Processing}

According to dual-process theories of cognition, information processing operates through two qualitatively distinct systems \citep{kahneman2011thinking}. System 1 operates automatically, rapidly, and with minimal conscious effort, relying on associative memory, heuristics, and pattern recognition to generate immediate intuitive judgments. System 2, in contrast, is slower, more deliberative, and analytical, engaging in controlled reasoning and conscious evaluation when situations require careful thought. Individuals rely predominantly on System 1 for most everyday judgments because it enables efficient processing of familiar information, reserving effortful System 2 engagement for novel situations or complex decisions. For bias measurement, this dual-process distinction has important implications. Research indicates that System 1 processing -- rapid, intuitive, and associative -- more faithfully reflects ingrained biases and automatic stereotypes, whereas System 2 processing enables deliberate correction and regulation of potentially biased initial responses \citep{devine1989stereotypes}. When given sufficient time and motivation, individuals can override automatic biases through conscious correction processes, particularly when egalitarian norms are salient. Consequently, measures that allow extensive deliberation may underestimate bias by capturing controlled, socially regulated responses rather than automatic perceptual tendencies.

To ensure that PictoPercept captures System 1 processing and reveals genuine spontaneous bias while minimizing conscious regulation, the task design incorporates relatively brief response windows of maximum five seconds per trial. This time constraint is sufficient to view both images and make a selection, but insufficient for extensive deliberation or conscious evaluation of how the choice might reflect on one's egalitarian self-concept. By limiting deliberation time, the design captures the immediacy of perceptual judgment, revealing unconscious and spontaneous biases as they naturally occur in rapid social perception. The brief presentation prevents participants from engaging in elaborate counterfactual reasoning or conscious self-monitoring. This approach builds on evidence from impression formation research demonstrating that initial face-based judgments form extremely rapidly -- within 100 milliseconds of exposure -- and that these immediate impressions strongly influence subsequent deliberative evaluations even when perceivers are given unlimited time \citep{willis2006first}. By capturing responses in a time frame that prioritizes intuitive System 1 processing while minimizing System 2 correction, PictoPercept reveals the perceptual biases that guide spontaneous social judgment in everyday contexts where people rarely have time for careful deliberation.


\subsubsection{Grounding Bias in Social Reality}

A crucial basis for the validity assessment of bias measures is the use of  external reference points. More often than not, both surveys and IATs are used to produce relative indices that reveal category-based preferences or associations, but they do not distinguish between accurate perception of real social inequalities and distorted perceptions that exaggerate, minimize, or misrepresent actual group differences. For example, since women do face barriers to career advancement in certain fields, is perceiving this disparity evidence of `bias' or simply accurate awareness of social reality? Conversely, if a perceiver believes that men and women have achieved full parity in professional domains where significant disparities persist, does this represent an absence of bias or rather a form of `colorblind' bias that underestimates ongoing inequality? PictoPercept addresses this conceptual challenge by grounding bias measurement in population-level benchmarks derived from authoritative data sources, such as U.S. Census Bureau statistics and Bureau of Labor Statistics reports \citep{us2021census, bls2025earnings} for the case presented here. Rather than simply measuring relative preferences between categories in the abstract, PictoPercept quantifies how much participants' perceptual judgments deviate from actual social distributions. This approach enables calculation of bias scores that reflect both direction (overestimation versus underestimation of group representation) and magnitude (how far perceptions deviate from reality).

Consider a concrete example: according to Bureau of Labor Statistics data, women constitute approximately 86.8\% of registered nurses in the United States while men represent 13.2\% \citep{bls2024nurses}. A participant using PictoPercept who selects female faces 87\% of the time when judging who is more likely to be a nurse would show minimal bias (accurate perception), whereas someone selecting female faces 95\% of the time would show overestimation of female dominance in nursing, and someone selecting female faces 50\% of the time would show dramatic underestimation of female representation. Traditional bias measures would identify only relative gender associations without reference to actual occupational demographics, potentially labelling accurate perceptions as `biased' or missing cases where apparently egalitarian beliefs actually represent substantial misperception of reality. This population-benchmarked approach offers several theoretical and practical advantages. First, it provides external validity by connecting perceptual measures to real-world social patterns, enabling researchers to assess whether bias operates to distort accurate perception or whether it reflects accuracy about existing inequalities. Second, it allows detection of bidirectional biases -- both stereotype exaggeration (believing inequalities are larger than they truly are) and system justification effects (minimizing real inequalities to maintain beliefs in meritocracy and fairness) \citep{jost2004decade}. Third, by anchoring measurements to objective data, the approach reduces ambiguity about what constitutes `bias' versus `accuracy,' providing clearer operational definitions for empirical research and potential intervention. Fourth, for AI systems, population-level benchmarking enables assessment of whether training data, model architectures, or fine-tuning procedures have introduced systematic distortions relative to ground truth, facilitating more precise diagnosis of fairness problems in machine learning pipelines. By grounding perceptual judgments in population-level reality rather than treating bias as purely relative or abstract, PictoPercept transforms comparative evaluations into interpretable, socially meaningful indicators of perceptual accuracy and distortion.

\section{Data and Methods}

\subsection{Overview and Participants}

Our study employs a mixed-methods online design combining experimental forced-choice tasks and self-report surveys to validate the PictoPercept bias measurement approach and assess earnings-related bias in both human participants and an AI system. Participants complete: (1) the PictoPercept task, consisting of 68 forced-choice trials presenting pairs of faces from the Chicago Face Database \citep{ma2015chicago}, a validated stimulus set with standardized photographs and normed ratings on multiple dimensions (see Figure 1 for an example); and (2) self-report measures including traditional text-based bias and social desirability surveys. Image pairs in the PictoPercept task are carefully balanced on non-focal covariates (attractiveness, trustworthiness, age) to isolate demographic effects from confounding perceptual features. After collecting human data, we further administer identical image pairs to GPT-5, a prominent AI vision-language model from OpenAI. For each participant (human or AI), we compute bias scores by comparing their selection patterns against U.S. Census and Bureau of Labor Statistics data on earnings across demographic groups, yielding both overall bias scores and dimension-specific scores for gender, ethnicity, and their interaction.

\begin{figure}[h!]
  \centering
  \includegraphics[width=0.7\linewidth]{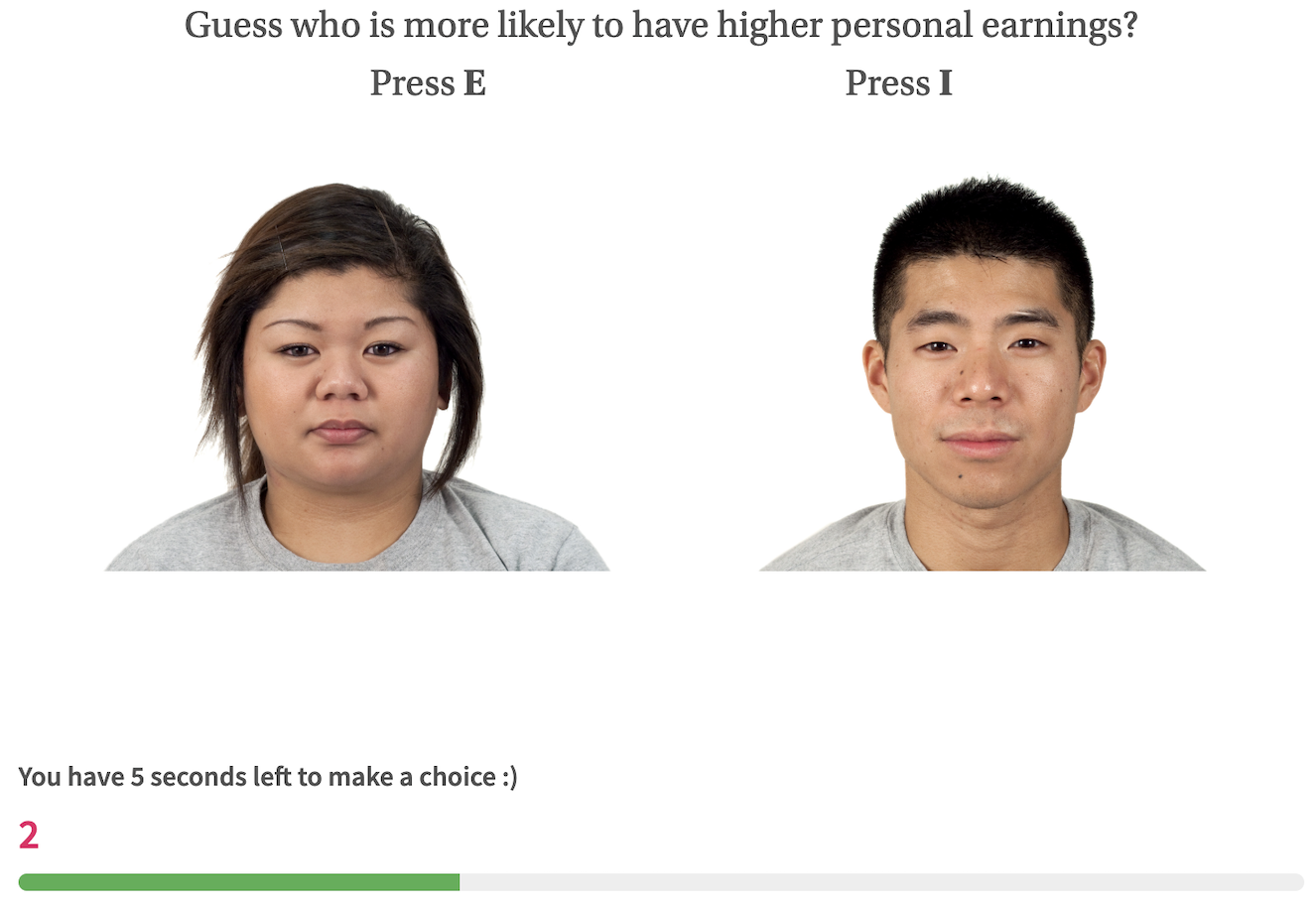}
  \caption{The example shows the \textit{PictoPercept} interface, where participants make a forced-choice judgment between two faces within a limited response time.}
  \label{fig:myimage}
\end{figure}

We recruited a nationally representative sample of 283 adult participants residing in the United States through Prolific Academic, a platform that enables recruitment of high-quality, diverse samples for online research \citep{palan2018prolific}. Participants received 1.8 GBP compensation for the 12 minute study. Following best practices in online research \citep{oppenheimer2009instructional}, during analysis we apply exclusion criteria including failure of more than one attention check, completion of fewer than 95\% of trials, or position bias exceeding 90\% (selecting the same screen location on more than 90\% of trials).

\subsection{Stimulus Materials and Procedure}

All facial photographs were drawn from the Chicago Face Database (CFD), a validated, standardized set of 597 high-resolution images representing diverse demographic characteristics \citep{ma2015chicago}. Each photograph depicts an individual with a neutral facial expression under standardized lighting conditions against a uniform gray background. Images are labeled with self-reported gender (Female, Male) and ethnicity (Asian, Black, White, Latino), and the database provides normed ratings on attractiveness, trustworthiness, dominance, age, and other attributes that enabled careful control for potential confounding variables when constructing image pairs.
Participants completed all study components remotely via Qualtrics for the PictoPercept task and self-report measures. After providing informed consent, participants completed three phases: (1) informed consent (1 minute), (2) PictoPercept image selection task (2--3 minutes), and (3) self-report measures (2--3 minutes).

\subsubsection{PictoPercept Image Selection Task}

The PictoPercept task constituted our primary measure of perceptual bias. Participants completed 68 forced-choice trials, each presenting two facial photographs side-by-side with the prompt: ``Which person is more likely to have higher personal earnings?'' Participants were instructed: ``There are no right or wrong answers -- we are interested in your initial impressions.'' This framing was designed to elicit intuitive, System 1 responses rather than deliberative reasoning \citep{kahneman2011thinking}. The 68 trials comprised 64 experimental trials and 4 attention check trials randomly interspersed. For experimental trials, image pairs were constructed to systematically vary across eight demographic categories (Male/Female $\times$ Asian/Black/Latino/White). Each demographic group appeared in exactly 16 trials, with careful counterbalancing across different pairings. Critically, image pairs were constructed to minimize differences on non-focal covariates: the two faces in each pair did not differ substantially on normed ratings of attractiveness, trustworthiness, dominance, perceived age, or median luminance (all differences $<$ 1 standard deviation on CFD normed scales). This matching procedure ensured that systematic selection patterns reflected responses to demographic characteristics rather than confounding perceptual features. For each trial, we recorded which image was selected, response timestamp, trial order, and demographic characteristics and covariate ratings for both faces. Image position (left versus right) was randomized across trials to prevent side biases.

\subsection{AI Model Assessment}

After completing human data collection, we assessed bias in GPT-5 (OpenAI), a prominent AI vision-language model. For the AI model, we presented the identical 64 experimental image pairs used in the human PictoPercept task with the same prompt. We used the model's standard API with default parameters (temperature = 0.7) to obtain responses. To ensure reliability, we ran the model through the full 64-trial sequence 100 times and used the majority response for each trial. In case the model provided explanatory text, we coded only the final selection to maintain comparability with human binary choice data. This approach enabled direct comparison between human and AI bias patterns using an identical measurement framework.

\subsection{Bias Score Calculation}

We calculated two types of bias scores for each participant (human or AI): Reality-Based Bias Scores (our primary outcome) and Ideal-Based Bias Scores (a secondary exploratory measure).

The Reality-Based Bias Score quantifies the extent to which participants' image selections deviate from patterns expected based on actual U.S. earnings data. We obtained median weekly earnings for full-time workers by gender and race/ethnicity from the U.S. Bureau of Labor Statistics, 2nd Quarter 2025 \citep{bls2025earnings}: Asian Male (\$1,759), Asian Female (\$1,363), White Male (\$1,357), White Female (\$1,100), Black Male (\$1,053), Black Female (\$942), Latino Male (\$1,005), and Latina Female (\$880). For each pairing of demographic groups, we calculated the expected selection probability based on relative earnings:

\begin{equation}
P_{\text{expected}}(i|i,j) = \frac{\text{Earnings}_i}{\text{Earnings}_i + \text{Earnings}_j}
\end{equation}

For example, when comparing an Asian Male face (earnings = \$1,759) with a Black Female face (earnings = \$942), perfect calibration to population reality would yield $P_{\text{expected}}(\text{Asian Male}) = \frac{1759}{1759+942} = 0.651$ or 65.1\%. For each participant, we calculated observed selection rates for each demographic group ($P_{\text{observed}}(i) = n_i/16$), then computed bias scores as:

\begin{equation}
\text{Bias}_{\text{reality}}(i) = P_{\text{observed}}(i) - P_{\text{expected}}(i)
\end{equation}

Positive scores indicate overestimation of a group's earnings relative to reality, while negative scores indicate underestimation. We calculated overall bias magnitude as mean absolute deviation across all eight groups:

\begin{equation}
\text{Overall Bias}_{\text{reality}} = \frac{1}{8}\sum_{i=1}^{8}|\text{Bias}_{\text{reality}}(i)|
\end{equation}

We also decomposed bias into dimensional components: \textit{Gender Bias} (difference between male and female selection rates averaged across ethnicities, compared to expected gender differences from earnings), \textit{Ethnicity Bias} (calculated similarly for each ethnic category), and \textit{Intersectional Bias} (examining whether bias for specific Gender $\times$ Ethnicity combinations exceeded additive predictions from main effects).

As an exploratory exercise, we also calculate an Ideal-Based Bias Score quantifies deviation from equality, assuming all demographic groups have identical earnings. The calculation parallels the Reality-Based score, except $P_{\text{expected, ideal}}(i|i,j) = 0.50$ for all pairings. Thus, $\text{Bias}_{\text{ideal}}(i) = P_{\text{observed}}(i) - 0.50$, with positive scores indicating systematic preference beyond parity and negative scores indicating systematic disadvantage. Reality-Based and Ideal-Based scores can diverge meaningfully: a participant who accurately perceives existing disparities would show low Reality-Based bias but potentially high Ideal-Based bias, whereas an AI model trained to ignore demographics entirely would show low Ideal-Based bias but high Reality-Based bias. Examining both metrics provides a more complete picture of perceptual patterns.

\section{Results}

A nationally representative sample of 283 American adults completed the PictoPercept task consisting of 64 experimental trials plus 4 attention check trials, making forced-choice judgments about which person was more likely to have higher personal earnings.
To assess the reliability of PictoPercept, we calculated bias scores separately for odd-numbered and even-numbered trials. The correlation between these two halves was $r = 0.637$ ($p < .001$), indicating moderate-to-good reliability. After applying the Spearman-Brown correction to estimate reliability for the full 64-trial measure, the adjusted reliability coefficient was $r_{sb} = 0.779$. This level of reliability is comparable to other implicit bias measures while requiring substantially shorter administration time.

\subsection{Earnings Perception Bias in Human Participants}

\subsubsection{Overall Patterns of Systematic Misperception}

\begin{figure}[h!]
  \centering
  \includegraphics[width=0.7\linewidth]{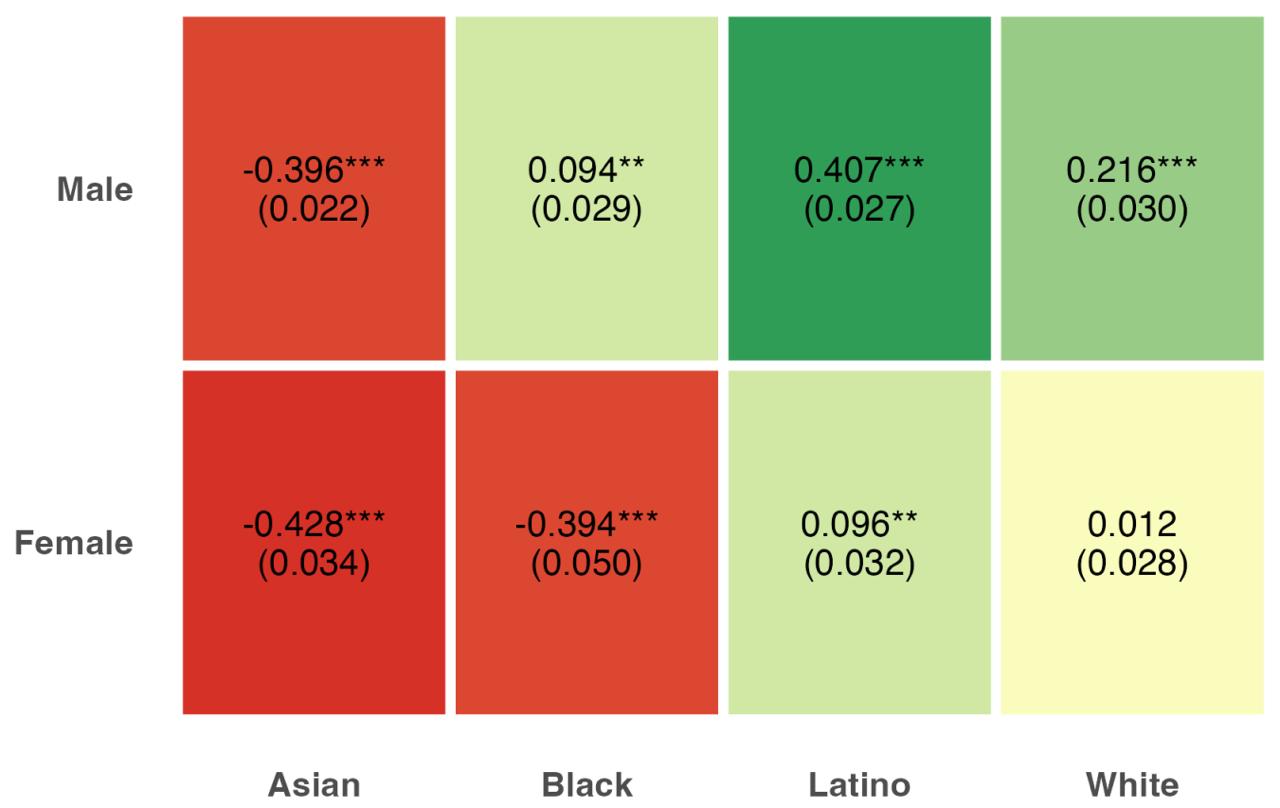}
  \caption{Earnings perception bias by facial image demographic group for all human participants. Values are means with standard errors in parentheses.}
  \label{fig:myimage2}
\end{figure}

Figure 2 presents the overall pattern of earnings perception bias across all eight demographic categories. Participants demonstrated systematic misperceptions that diverged substantially from actual U.S. earnings data. Negative bias scores indicate underestimation (selecting a group less frequently than their actual earnings would predict), while positive scores indicate overestimation.
The most striking finding was the severe underestimation of Asian American earnings. Despite Asian Americans representing the highest-earning demographic group in U.S. labor statistics, participants dramatically underestimated both Asian male ($M = -0.396$, $SE = 0.022$, $p < .001$) and Asian female earnings ($M = -0.428$, $SE = 0.034$, $p < .001$). Black females were similarly underestimated ($M = -0.394$, $SE = 0.050$, $p < .001$), while Black males showed modest overestimation.
In contrast, participants substantially overestimated earnings for Latino males ($M = 0.407$, $SE = 0.027$, $p < .001$) and White males ($M = 0.216$, $SE = 0.030$, $p < .001$), selecting these groups far more frequently than their actual earnings justified. Latino females and White females showed more accurate perceptions, with bias scores closer to zero. This pattern suggests that participants hold systematic stereotypes that diverge markedly from economic reality, particularly for Asian Americans and Latino men.

\subsubsection{Gender Differences in Perception}

\begin{figure}[h!]
  \centering
  \includegraphics[width=0.7\linewidth]{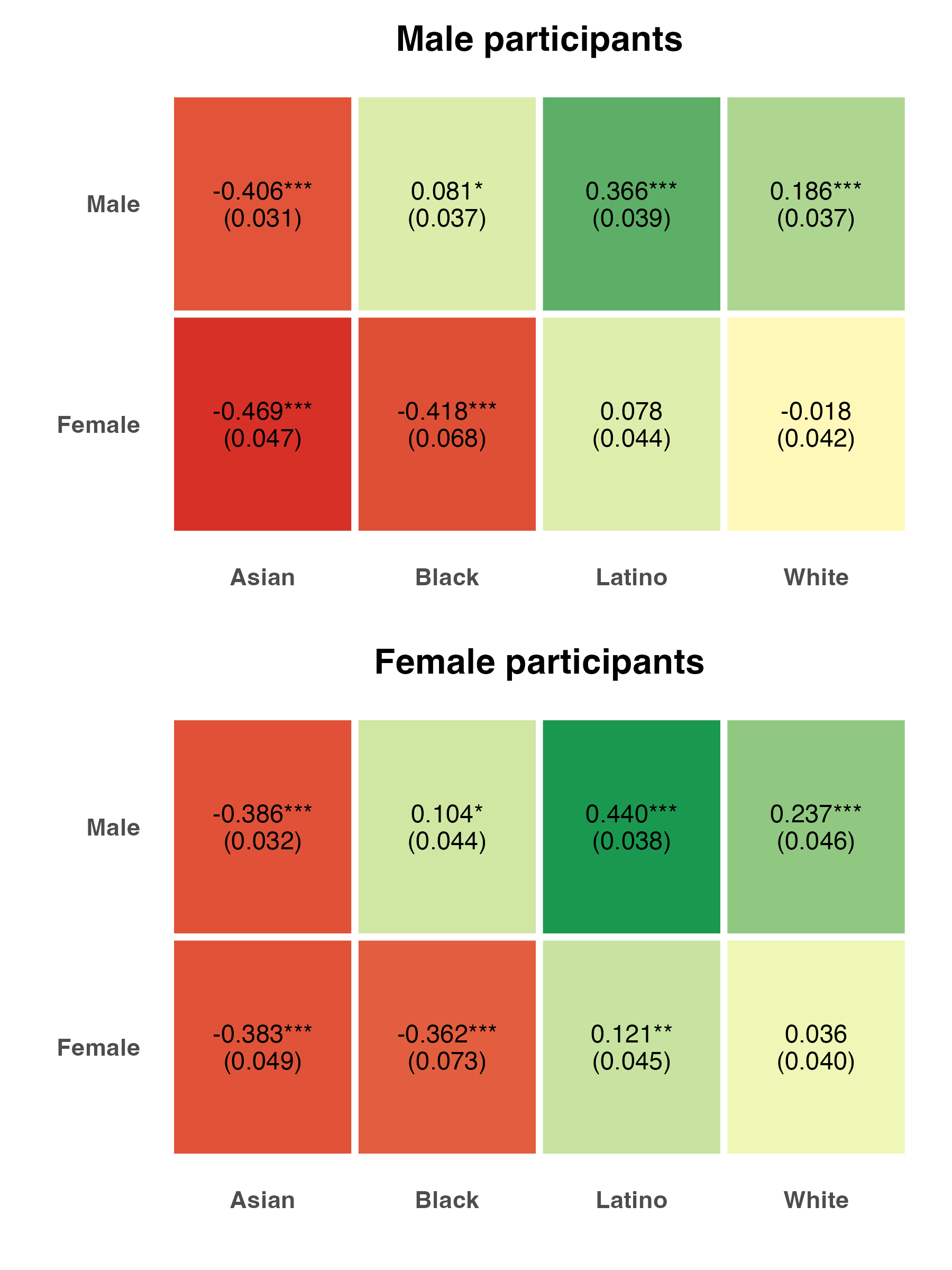}
  \caption{Earnings perception bias for male and female participants by demographic group of facial images. Values are means with standard errors in parentheses.}
  \label{fig:myimage3}
\end{figure}

Male and female participants exhibited similar directional biases but differed substantially in magnitude (Figure 3). Both groups underestimated Asian and Black female earnings while overestimating Latino and White male earnings. We find that female participants showed stronger overestimation of Latino male earnings ($M = 0.440$, $SE = 0.038$) compared to male participants ($M = 0.366$, $SE = 0.039$). On the other hand, male participants showed more extreme underestimation of Asian females ($M = -0.469$, $SE = 0.047$ for males vs. $M = -0.383$, $SE = 0.049$ for females) and Black females.
Male participants displayed similar bias magnitudes overall, with several demographic categories showing near-accurate perceptions (e.g., White females, Latino females). Overall we find both genders to share similar stereotypical associations.

\subsubsection{The Role of Participants' Own Ethnicity}

\begin{figure}[h!]
  \centering
  \includegraphics[width=\linewidth]{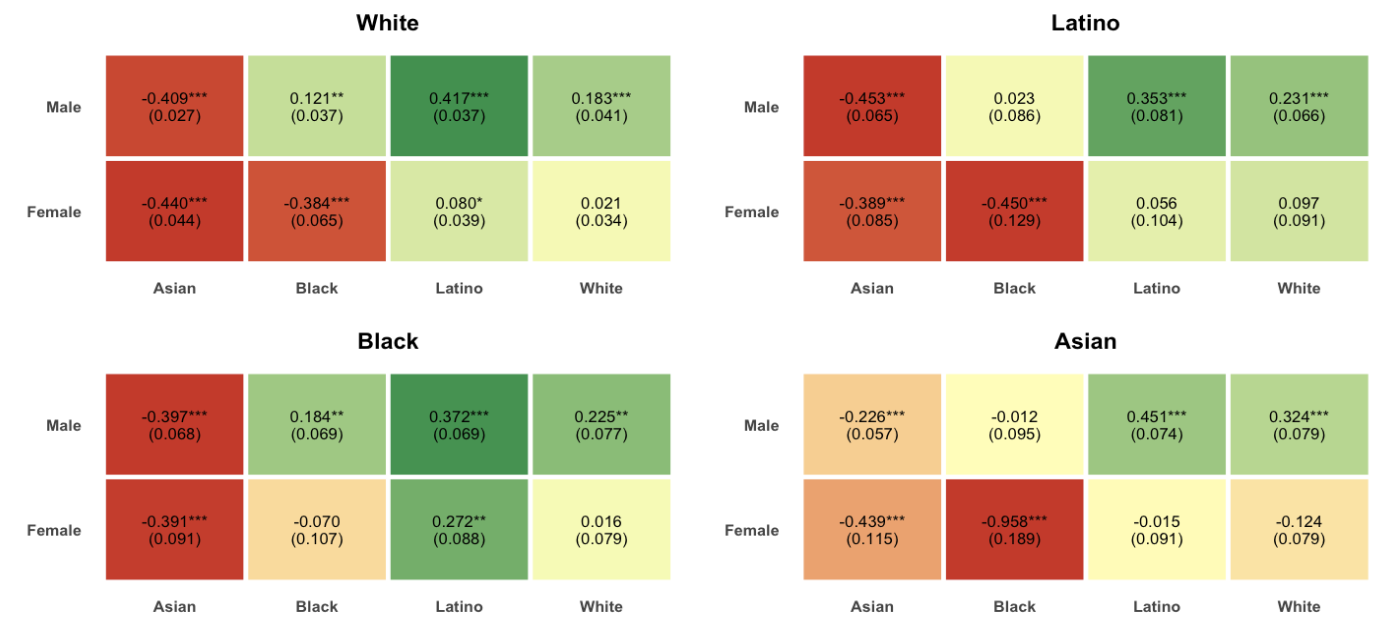}
  \caption{Earnings perception bias by demographic group for participants of different ethnicities. Values are means with standard errors in parentheses.}
  \label{fig:myimage4}
\end{figure}

We examined whether participants' own ethnicity shaped their perception patterns (Figure 4). The core pattern of underestimating Asian earnings and overestimating Latino male earnings emerged consistently across all ethnic groups, suggesting these stereotypes are widely shared across American society.
However, important differences emerged. White participants showed the expected pattern of bias across all groups. Black participants showed less bias when judging their own racial group -- Black female earnings were not significantly underestimated ($M = -0.070$, $p = .51$), while other groups showed strong biases consistent with the overall sample. Latino participants similarly showed non-significant bias for Latino females but maintained strong overestimation of Latino males.

Most strikingly, Asian participants showed substantial negative ingroup bias, significantly underestimating both Asian male ($M = -0.226$, $SE = 0.057$, $p < .001$) and Asian female earnings ($M = -0.439$, $SE = 0.115$, $p < .001$). Asian participants were also the only group to show extreme underestimation of Black females ($M = -0.958$, $SE = 0.189$, $p < .001$). These findings suggest that awareness of Asian American economic success is limited even within Asian American communities, potentially reflecting the relative invisibility of this demographic pattern in mainstream discourse.

\subsection{Ingroup Favoritism}

\begin{figure}[h!]
  \centering
  \includegraphics[width=\linewidth]{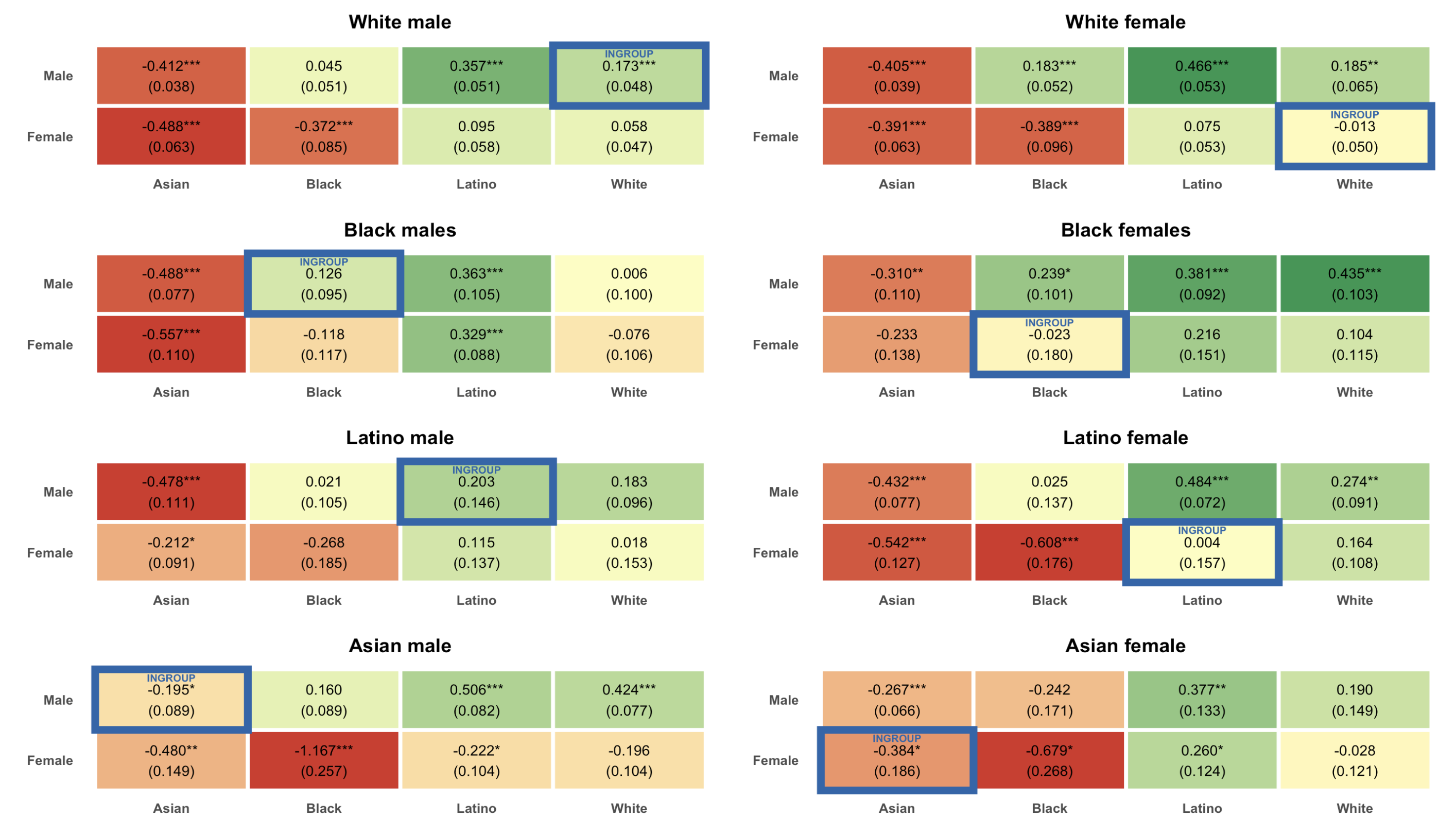}
  \caption{Earnings perception bias by demographic group. Values are means with standard errors in parentheses.}
  \label{fig:myimage5}
\end{figure}

A central question in bias research concerns whether people systematically favor their own demographic groups. Our results revealed striking heterogeneity in ingroup bias patterns (Figure 5). White male participants exhibited clear ingroup favoritism, significantly overestimating White male earnings ($M = 0.173$, $SE = 0.048$, $p < .001$). In contrast, White female participants showed no ingroup bias whatsoever ($M = -0.013$, $p = .79$), perceiving White female earnings with near-perfect accuracy.
Black male and Black female participants showed minimal ingroup bias, with bias scores near zero and non-significant. Latino participants similarly showed accurate perceptions of their own gender-ethnicity combinations. The most remarkable finding emerged among Asian participants, who demonstrated significant \textit{negative} ingroup bias. Asian males underestimated Asian male earnings ($M = -0.195$, $p < .05$), and Asian females showed even stronger underestimation of their own group ($M = -0.384$, $p < .05$).

These patterns challenge the assumption that ingroup favoritism is a universal feature of social perception. While some groups (particularly White males) show clear ingroup preference, other groups either perceive their ingroup accurately or actually underestimate their ingroup's economic position. This heterogeneity may reflect differences in group visibility, media representation, or the salience of economic disparities within different communities.

\subsection{Moderating Effects of Individual Differences}

\subsubsection{Political Ideology}

\begin{figure}[h!]
  \centering
  \includegraphics[width=\linewidth]{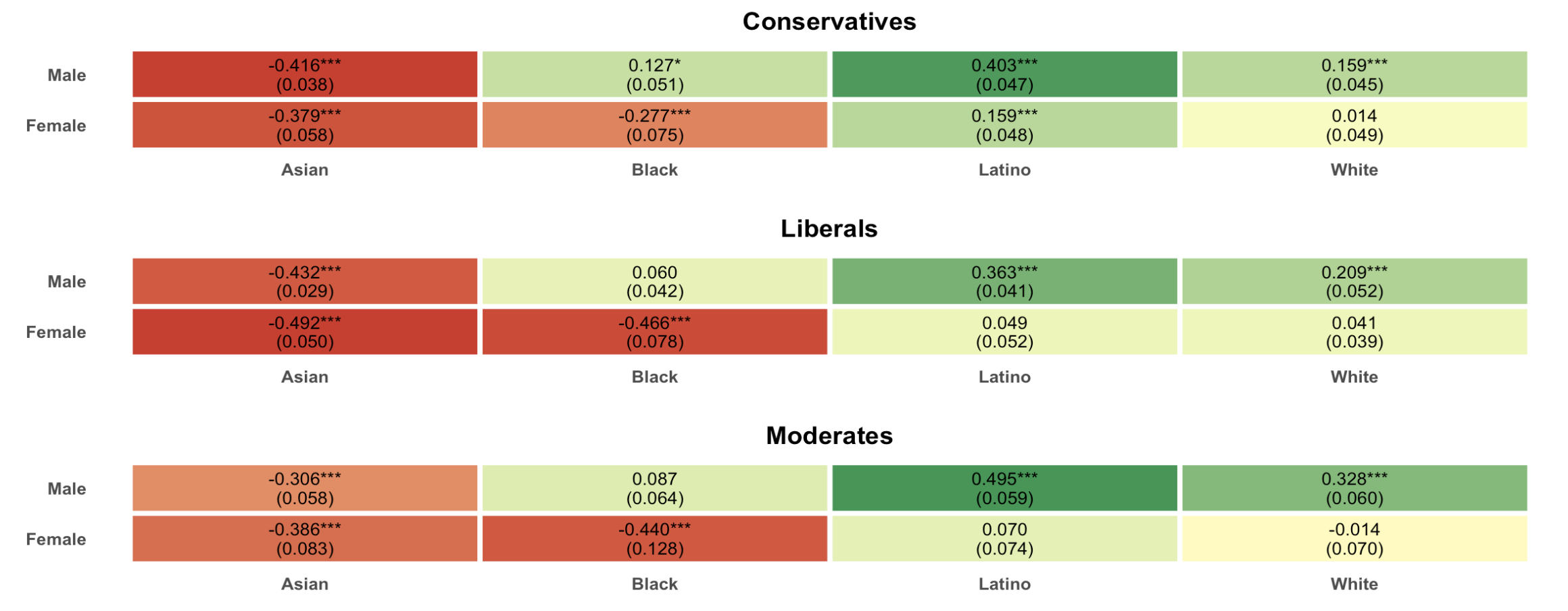}
  \caption{Earnings perception bias by facial image demographic group for conservation, liberal, moderate participants. Values are means with standard errors in parentheses.}
  \label{fig:myimage6}
\end{figure}

Political ideology moderated the magnitude but not the overall direction of bias (Figure 6). Conservative, moderate, and liberal participants all showed the core pattern of underestimating Asian earnings and overestimating Latino male earnings. However, moderates showed the strongest overestimation of Latino males ($M = 0.495$, $p < .001$), while conservatives and liberals showed somewhat smaller but still substantial biases in the same direction. Liberals showed the strongest underestimation of Asian females ($M = -0.492$, $p < .001$), suggesting that progressive political orientation does not necessarily correspond to more accurate perceptions of economic inequality across demographic groups.

\subsubsection{Socioeconomic Status}

\begin{figure}[h!]
  \centering
  \includegraphics[width=.8\linewidth]{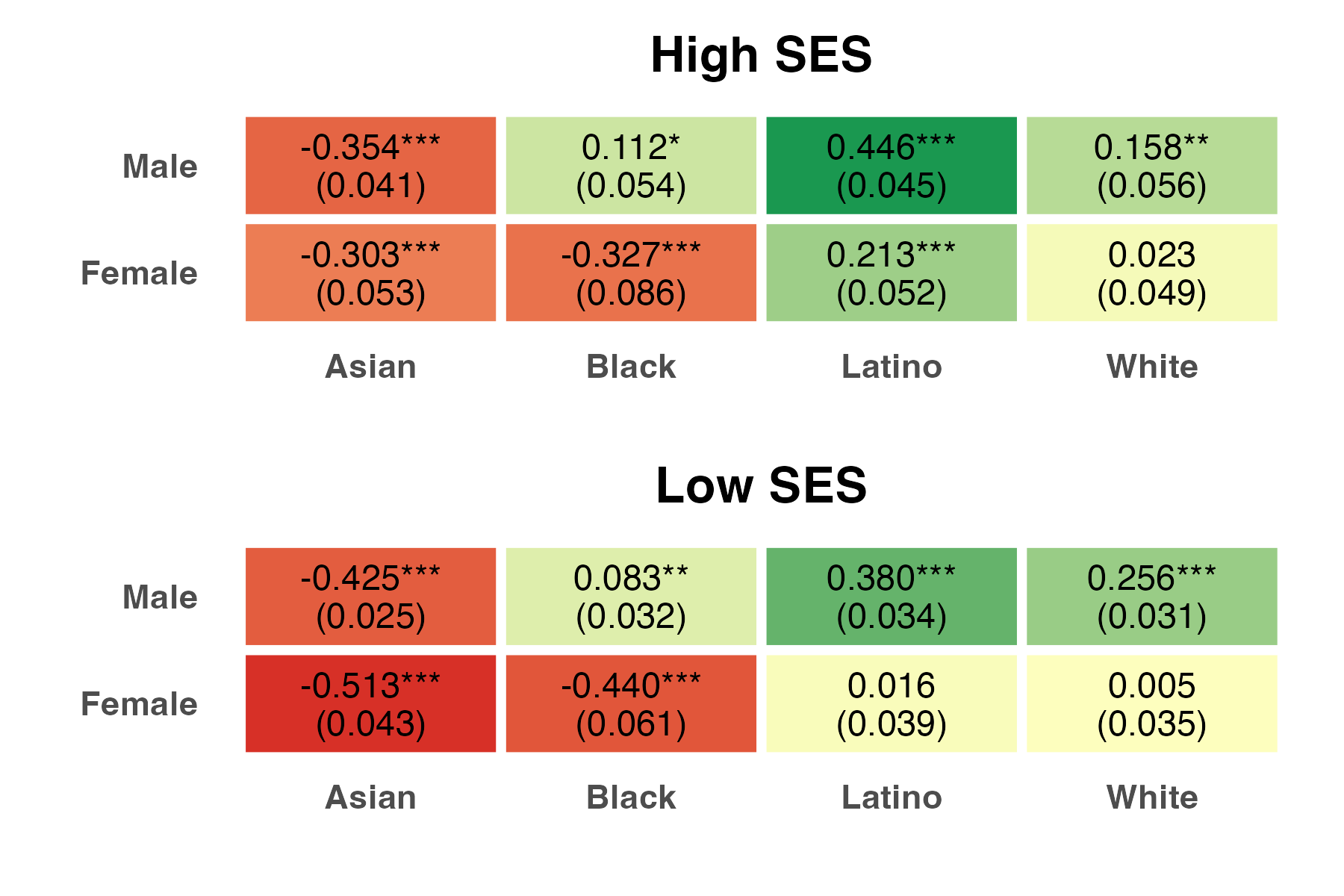}
  \caption{Earnings perception bias by demographic group. Values are means with standard errors in parentheses.}
  \label{fig:myimage7}
\end{figure}

Participants' self-reported socioeconomic status showed modest moderating effects (Figure 7). Both high-SES and low-SES participants demonstrated the core bias patterns, though low-SES participants showed somewhat stronger underestimation of Asian females ($M = -0.513$, $p < .001$) compared to high-SES participants ($M = -0.303$, $p < .001$). High-SES participants showed stronger overestimation of Latino females ($M = 0.213$, $p < .001$) compared to low-SES participants ($M = 0.016$, $p = .68$). These patterns suggest that direct economic experience may not substantially improve the accuracy of earnings perceptions across demographic groups.

\subsubsection{Geographic Location}

\begin{figure}[h!]
  \centering
  \includegraphics[width=\linewidth]{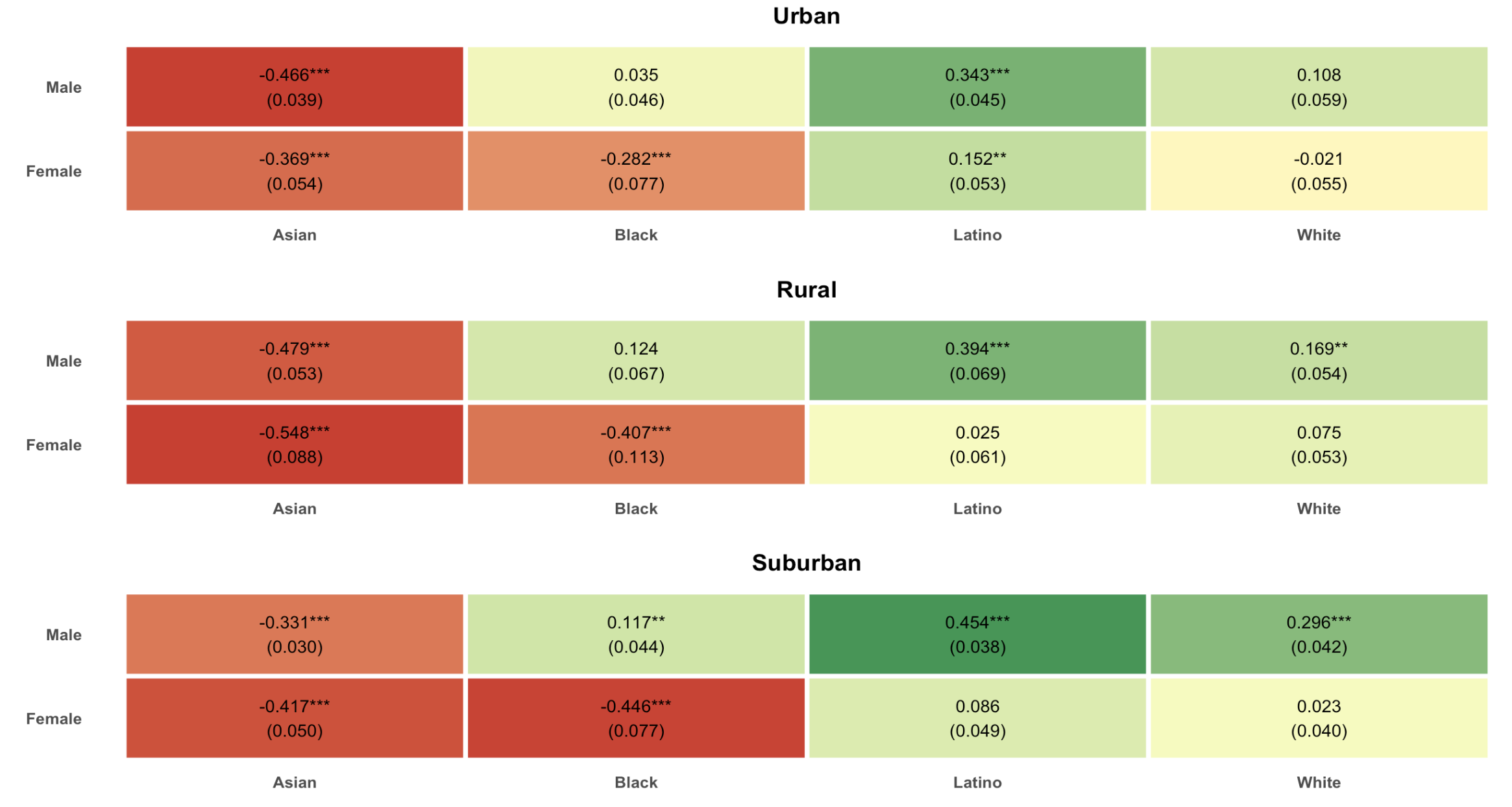}
  \caption{Earnings perception bias by demographic group. Values are means with standard errors in parentheses.}
  \label{fig:myimage8}
\end{figure}

Geographic differences revealed more substantial variation. Urban participants showed more moderate biases overall, with non-significant bias for several groups including White males and White females. In contrast, rural and suburban participants showed stronger and more consistent bias patterns across most demographic categories. Rural participants showed the strongest underestimation of Asian females ($M = -0.548$, $p < .001$), while suburban participants showed the strongest overestimation of White males ($M = 0.296$, $p < .001$). These geographic differences may reflect differential exposure to demographic diversity or different local labor market compositions.

\subsection{AI Bias}

\begin{figure}[h!]
  \centering
  \includegraphics[width=.7\linewidth]{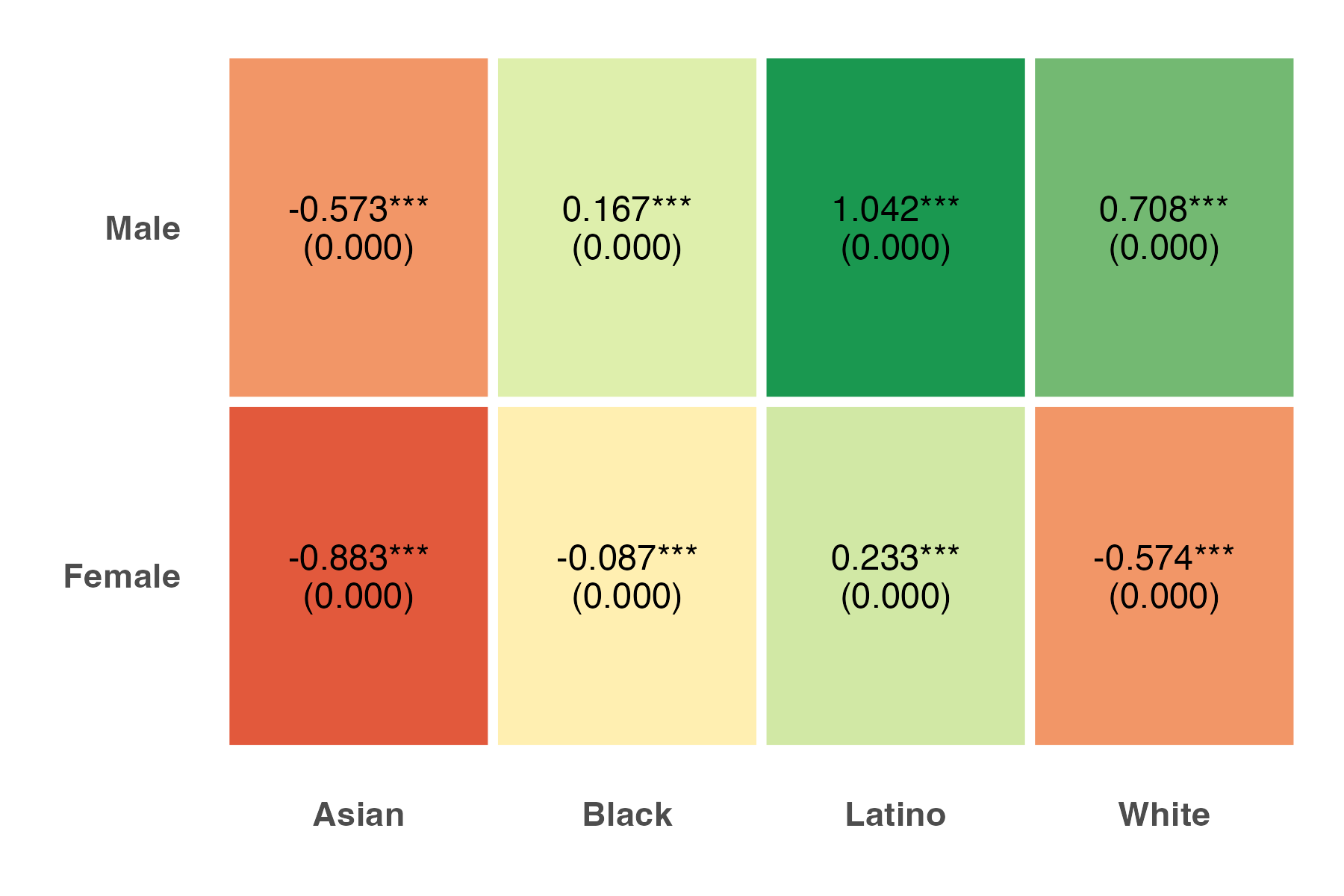}
  \caption{GPT-5's earnings perception bias by facial image demographic group. Values are means with standard errors in parentheses.}
  \label{fig:myimage9}
\end{figure}

We administered the identical PictoPercept task to GPT-5 using 100 repetitions to ensure stability. The AI model's bias is largely consistent with the patterns observed in human averages, especially in terms of gender bias (Figure 9). 
GPT-5 severely underestimated earnings for most of female groups: Asian females ($M = -0.883$), White females ($M = -0.574$), and Black females ($M = -0.087$). Conversely, the model dramatically overestimated earnings for most male groups, with Latino males showing the most extreme overestimation ($M = 1.042$), followed by White males ($M = 0.708$) and Black males ($M = 0.167$).

However, the magnitude of GPT-5's biases substantially exceeded human averages. This amplification of bias -- particularly along the gender dimension -- raises significant concerns about deploying vision-language AI systems in consequential decision-making contexts such as hiring, lending, or performance evaluation. The AI's extreme gender bias suggests that even state-of-the-art models may encode and amplify societal stereotypes present in training data, potentially creating more biased decisions than human evaluators would make.

In summary, PictoPercept revealed systematic earnings perception biases across all demographic groups, with three key findings: First, participants dramatically underestimated Asian American earnings despite this group having the highest actual earnings, while overestimating Latino male and White male earnings. Second, ingroup favoritism was not universal -- White males showed clear ingroup bias, but Asian participants actually underestimated their own group's earnings. Third, GPT-5 exhibited substantially stronger biases than humans, with extreme systematic underestimation of all female groups and overestimation of male groups. These patterns held across diverse subgroups, though with varying magnitudes based on participant characteristics.

\section{Discussion}

This study introduced and validated PictoPercept, a novel open-source toolkit for measuring bias in both human and artificial intelligence systems through visual forced-choice judgments. Our findings reveal systematic earnings perception biases that diverge substantially from demographic reality, challenge traditional assumptions about ingroup favoritism, and demonstrate that state-of-the-art AI systems amplify rather than attenuate human stereotypes. We discuss the theoretical and practical implications of these findings below.

Perhaps the most striking finding from our study is the severe and consistent underestimation of Asian American earnings across all participant groups. Despite Asian Americans representing the highest-earning demographic in U.S. labor statistics -- with Asian males earning \$1,759 and Asian females earning \$1,363 per week compared to national averages -- participants dramatically underestimated their economic success. This pattern held across participant demographics, political ideologies, and socioeconomic backgrounds, suggesting a widely shared perceptual blind spot.
This systematic underestimation is particularly puzzling given the ``model minority'' stereotype often associated with Asian Americans. Traditional stereotype research has documented perceptions of Asian Americans as high-achieving and academically successful, yet our findings suggest these associations do not translate into accurate perceptions of economic outcomes. Several theoretical mechanisms may explain this paradox. First, the aggregation of diverse Asian ethnic groups (East Asian, South Asian, Southeast Asian) into a single category may obscure economic variation within this population, with some subgroups facing substantial economic challenges. Second, the relative numerical minority status of Asian Americans may reduce their visibility in leadership and high-earning positions despite statistical overrepresentation. Third, cultural stereotypes emphasizing technical competence over social influence may lead perceivers to associate Asian Americans with mid-level professional roles rather than executive positions commanding the highest salaries.
The fact that even Asian participants dramatically underestimated their own group's earnings -- demonstrating significant negative ingroup bias -- suggests that awareness of Asian American economic success is limited even within these communities. This internal misperception may reflect the broader invisibility of Asian American representation in popular media, corporate leadership portrayals, and public discourse about economic inequality. When economic disparities are discussed, Asian Americans are often excluded from analyses or subsumed into broader categories, potentially contributing to collective unawareness of this group's economic position.

Classical social identity theory posits that individuals systematically favor their ingroups across evaluative domains. Our findings challenge this assumption, revealing striking heterogeneity in ingroup bias patterns across demographic groups. White male participants exhibited clear ingroup favoritism, significantly overestimating White male earnings relative to reality. In contrast, White female participants showed near-perfect accuracy when judging their own group, demonstrating no ingroup bias whatsoever. Black and Latino participants showed minimal ingroup bias, while Asian participants demonstrated the opposite pattern -- significant negative ingroup bias.
This heterogeneity suggests that ingroup favoritism is not an automatic feature of social cognition but rather depends on contextual factors including group visibility, status, and the salience of existing inequalities. White males, occupying a historically dominant position in economic hierarchies, may feel entitled to overestimate their group's success or may accurately perceive historical advantages while misattributing them to individual merit. In contrast, groups facing documented economic barriers may internalize societal narratives about their disadvantage, leading to accurate or even pessimistic perceptions of their economic position.
The negative ingroup bias observed among Asian participants is particularly theoretically important. It suggests that the same mechanisms that render Asian American success invisible to outgroup members also affect ingroup perceptions. This finding has implications beyond academic interest: if members of economically successful groups fail to recognize their collective advantages, they may be less likely to advocate for policies addressing the needs of less advantaged subgroups within their broader community or to challenge structural barriers facing other marginalized groups.

While the core pattern of underestimating Asian earnings and overestimating Latino male earnings appeared consistently across subgroups, the magnitude of these biases varied systematically. Female participants demonstrated stronger biases than male participants across most demographic categories. This gender difference may reflect differential exposure to earnings information, with women potentially less likely to occupy positions providing direct visibility into compensation patterns, or alternatively, women may hold more polarized stereotypes as a result of their own experiences with gender-based economic disadvantage.
Political ideology moderated bias magnitude but not direction, with conservatives, moderates, and liberals all showing similar stereotypical patterns. This finding challenges the assumption that progressive political orientation necessarily corresponds to more accurate perceptions of demographic inequality. Indeed, liberals showed the strongest underestimation of Asian female earnings, suggesting that ideological commitment to egalitarian principles does not automatically translate into empirically grounded understanding of existing disparities. This dissociation between political values and perceptual accuracy has important implications for policy discourse: well-intentioned efforts to address inequality may be misdirected if they rest on inaccurate assessments of which groups face the most severe disadvantages.
Geographic differences revealed the largest moderating effects, with urban participants showing substantially more accurate perceptions than suburban and rural participants. These differences likely reflect greater demographic diversity and direct exposure to varied economic outcomes in urban environments. However, even urban participants demonstrated significant biases, suggesting that proximity alone does not eliminate stereotypical thinking. These findings underscore the importance of not only increasing diversity in physical spaces but also ensuring visibility and recognition of diverse economic achievements.

Our assessment of GPT-5 revealed bias patterns that substantially exceeded human averages in both magnitude and consistency. While human participants showed complex patterns reflecting both gender and ethnicity, the AI model exhibited an overwhelmingly dominant gender effect, severely underestimating all female earnings while dramatically overestimating male earnings. Latino males received the most extreme overestimation, with bias scores exceeding 1.0 -- more than double the typical human bias magnitude.
This AI bias amplification has several potential sources. Vision-language models are trained on vast corpora of internet text and images that encode existing societal stereotypes and may overrepresent historical periods or contexts where gender disparities were more pronounced. The model's training objective -- predicting patterns in data -- may lead it to exaggerate statistical regularities rather than accurately represent current demographic distributions. Additionally, the absence of deliberative reasoning or explicit calibration to population statistics means the model has no mechanism for correcting stereotypical associations that emerge from its training data.
The practical implications are sobering. As AI systems are increasingly deployed in consequential domains including hiring, performance evaluation, lending, and criminal justice, our findings suggest that these systems may produce more biased decisions than human evaluators operating under similar conditions. While human bias is substantial and concerning, AI bias is systematic, reproducible, and operates at scale without the possibility of individual reflection or correction. Organizations deploying such systems must implement rigorous auditing procedures and consider whether algorithmic decision-making actually improves upon human judgment or merely automates and amplifies existing inequalities.

Beyond substantive findings, this study demonstrates several methodological advantages of the PictoPercept approach. First, the measure achieved adequate reliability while requiring substantially less administration time than traditional IAT protocols. The split-half reliability is comparable to established implicit measures, suggesting the value of PictoPercept as a psychometrically sound instrument.
Second, the visual forced-choice format successfully reduced social desirability effects by embedding bias assessment within ostensibly neutral comparative judgments rather than explicit category evaluations. Participants made rapid, intuitive selections without consciously regulating their responses according to egalitarian norms, allowing genuine perceptual biases to emerge. The comparative format also enabled simultaneous assessment across multiple demographic dimensions without requiring separate testing blocks for each category or combination, addressing a fundamental limitation of traditional factorial designs.
Third, grounding bias measurement in population-level benchmarks rather than relative preferences transformed abstract bias indices into interpretable indicators of perceptual accuracy and distortion. This approach allowed us to distinguish between accurate awareness of existing inequality and biased exaggeration or minimization of disparities -- a distinction impossible with measures producing only relative preference scores. The same measurement framework applied equally to human and AI systems, enabling direct comparison and revealing that AI bias patterns differ qualitatively from human stereotypes.
Fourth, the open-source nature of PictoPercept promotes reproducibility and enables researchers to adapt the tool for different domains (e.g., occupation stereotypes, leadership perceptions, healthcare disparities) or populations. Unlike proprietary measures, PictoPercept can be freely modified, extended, and validated across contexts, facilitating cumulative knowledge development.

Several limitations warrant acknowledgment. First, our study focused exclusively on earnings-related bias using a single question prompt. While earnings represent a consequential domain where bias has material implications, future research should extend PictoPercept to other judgment domains including hiring decisions, competence evaluations, leadership potential, and healthcare treatment recommendations. Different domains may elicit different stereotypes, and the generalizability of our findings across contexts remains an empirical question.
Second, our stimulus set was limited to the Chicago Face Database, which provides standardized photographs but necessarily constrains the range of facial features, expressions, and contextual cues available. While we carefully controlled for confounding variables including attractiveness and trustworthiness, other perceptual features not captured in CFD norming data may influence judgments. Future work should replicate these findings with alternative stimulus sets and explore whether bias magnitudes vary with specific facial characteristics beyond broad demographic categories.
Third, our sample, while nationally representative, was limited to U.S. adults. Earnings stereotypes likely vary across national contexts with different demographic compositions, economic structures, and cultural narratives about success. Cross-national validation would establish the extent to which our findings reflect universal stereotype content versus culturally specific beliefs. Additionally, our reliance on self-reported demographic categories may obscure within-group heterogeneity, particularly for broad categories like ``Asian'' or ``Latino'' that encompass diverse ethnic and national origins.
Fourth, our AI assessment included only GPT-5, and while this represents a state-of-the-art vision-language model, bias patterns may vary across models with different architectures, training data, or fine-tuning procedures. Systematic comparison across multiple AI systems would clarify whether our findings reflect general properties of vision-language models or idiosyncrasies of particular implementations. Future research should also investigate potential debiasing interventions for AI systems, testing whether explicit calibration to population statistics or adversarial training procedures can reduce stereotypical associations.
Fifth, while PictoPercept captures perceptual biases, it does not directly assess consequential discrimination -- the behavioral translation of stereotypical beliefs into differential treatment. The relationship between measured bias and discriminatory behavior is complex, mediated by situational factors, accountability concerns, and individual motivations. Validating PictoPercept against behavioral outcomes in realistic decision contexts would strengthen claims about its predictive validity for real-world inequality.
Finally, our study was cross-sectional, providing a snapshot of bias patterns at a single time point. Tracking how these perceptions evolve as demographic compositions and economic disparities change would reveal whether stereotypes lag behind reality or adapt dynamically to shifting social conditions. Longitudinal assessment could also evaluate whether interventions designed to increase awareness of demographic economic patterns successfully reduce perceptual biases.

Our findings carry several practical implications for organizations, policymakers, and educators seeking to address bias and promote equity. First, the severe underestimation of Asian American earnings suggests that diversity initiatives focusing exclusively on representation may overlook groups whose economic success is already substantial but inadequately recognized. Increasing awareness of Asian American achievement -- while simultaneously attending to within-group heterogeneity and challenges facing specific subgroups -- could help counter invisibility and ensure equitable recognition.
Second, the overestimation of Latino male earnings suggests that compensation decisions may be influenced by inflated assumptions about this group's economic position. Organizations should implement transparent compensation structures and regularly audit salary decisions to ensure that stereotypical beliefs do not translate into unjustified wage premiums or penalties. Training programs that present actual demographic earnings data may help calibrate perceptions and reduce stereotype-driven decisions.
Third, the extreme AI bias we documented argues strongly for mandatory algorithmic auditing before deploying vision-language models in high-stakes domains. Organizations should not assume that AI systems provide unbiased alternatives to human judgment; indeed, our findings suggest that algorithmic decisions may be substantially more biased than human evaluations. Regulatory frameworks should require transparency about model training data, validation against population benchmarks, and ongoing monitoring for disparate impacts.
Fourth, the heterogeneity in ingroup bias patterns suggests that interventions promoting ingroup pride or collective identity may have different effects across groups. While such interventions might benefit groups showing negative ingroup bias (such as Asian Americans), they could exacerbate existing favoritism among groups already overestimating their achievements (such as White males). Tailored approaches sensitive to specific group experiences may be more effective than universal diversity programming.

\subsection{Conclusion}

Our study introduced \textit{PictoPercept}, an open-source (available on \href{https://github.com/theinvisiblelab/pictopercept}{GitHub}) toolkit enabling unified assessment of bias in humans and AI systems through visual forced-choice judgments grounded in population reality. Our findings revealed systematic misperceptions of earnings across demographic groups, with particularly severe underestimation of Asian American economic success despite their position as the highest-earning group. Ingroup favoritism emerged as heterogeneous rather than universal, with some groups accurately perceiving their economic position while others demonstrated negative ingroup bias. Most alarmingly, GPT-5 exhibited substantially stronger biases than human averages, with extreme gender-based distortions raising serious concerns about deploying vision-language AI in consequential domains.

These findings carry both theoretical and practical significance. Theoretically, they demonstrate that bias measurement benefits from visual rather than textual stimuli, simultaneous rather than sequential dimension assessment, and grounding in population benchmarks rather than relative preferences. Substantively, they reveal that stereotypes about economic success do not necessarily align with demographic reality, that ingroup favoritism depends on group status and visibility, and that AI systems may amplify rather than attenuate human biases. Practically, they argue for greater attention to the invisibility of Asian American achievement, systematic auditing of AI systems before deployment in high-stakes contexts, and recognition that ideological commitment to equality does not automatically translate into accurate perceptions of existing disparities.

As societies become increasingly diverse and as AI systems assume greater decision-making authority, tools enabling rigorous assessment of bias across both human and machine intelligence become essential. PictoPercept offers one such tool -- open, validated, and applicable across domains. Our hope is that researchers, practitioners, and policymakers will adapt and extend this approach to promote fairness, accountability, and empirically grounded understanding of how bias shapes consequential social and economic outcomes.

\onehalfspacing

\bibliographystyle{apalike}
\bibliography{references}

@article{mehrabi2021survey,
  title={A survey on bias and fairness in machine learning},
  author={Mehrabi, Ninareh and Morstatter, Fred and Saxena, Nripsuta and Lerman, Kristina and Galstyan, Aram},
  journal={ACM computing surveys (CSUR)},
  volume={54},
  number={6},
  pages={1--35},
  year={2021},
  publisher={ACM New York, NY, USA}
}

@article{gawronski2006associative,
  title={Associative and propositional processes in evaluation: an integrative review of implicit and explicit attitude change.},
  author={Gawronski, Bertram and Bodenhausen, Galen V},
  journal={Psychological bulletin},
  volume={132},
  number={5},
  pages={692},
  year={2006},
  publisher={American Psychological Association}
}

@article{greenwald1998measuring,
  title={Measuring individual differences in implicit cognition: the implicit association test.},
  author={Greenwald, Anthony G and McGhee, Debbie E and Schwartz, Jordan LK},
  journal={Journal of personality and social psychology},
  volume={74},
  number={6},
  pages={1464},
  year={1998},
  publisher={American Psychological Association}
}

@book{schuman1996questions,
  title={Questions and answers in attitude surveys: Experiments on question form, wording, and context},
  author={Schuman, Howard and Presser, Stanley},
  year={1996},
  publisher={Sage}
}

@article{crowne1960new,
  title={A new scale of social desirability independent of psychopathology.},
  author={Crowne, Douglas P and Marlowe, David},
  journal={Journal of consulting psychology},
  volume={24},
  number={4},
  pages={349},
  year={1960},
  publisher={American Psychological Association}
}

@article{greenwald2009understanding,
  title={Understanding and using the Implicit Association Test: III. Meta-analysis of predictive validity.},
  author={Greenwald, Anthony G and Poehlman, T Andrew and Uhlmann, Eric Luis and Banaji, Mahzarin R},
  journal={Journal of personality and social psychology},
  volume={97},
  number={1},
  pages={17},
  year={2009},
  publisher={American Psychological Association}
}

@article{hofmann2005meta,
  title={A meta-analysis on the correlation between the Implicit Association Test and explicit self-report measures},
  author={Hofmann, Wilhelm and Gawronski, Bertram and Gschwendner, Tobias and Le, Huy and Schmitt, Manfred},
  journal={Personality and social psychology bulletin},
  volume={31},
  number={10},
  pages={1369--1385},
  year={2005},
  publisher={Sage Publications Sage CA: Thousand Oaks, CA}
}

@article{ghavami2013intersectional,
  title={An intersectional analysis of gender and ethnic stereotypes: Testing three hypotheses},
  author={Ghavami, Negin and Peplau, Letitia Anne},
  journal={Psychology of Women Quarterly},
  volume={37},
  number={1},
  pages={113--127},
  year={2013},
  publisher={Sage Publications Sage CA: Los Angeles, CA}
}

@article{blasi2006system,
  title={System justification theory and research: Implications for law, legal advocacy, and social justice},
  author={Blasi, Gary and Jost, John T},
  journal={California Law Review},
  volume={94},
  number={4},
  pages={1119--1168},
  year={2006},
  publisher={JSTOR}
}

@inproceedings{friedler2019comparative,
  title={A comparative study of fairness-enhancing interventions in machine learning},
  author={Friedler, Sorelle A and Scheidegger, Carlos and Venkatasubramanian, Suresh and Choudhary, Sonam and Hamilton, Evan P and Roth, Derek},
  booktitle={Proceedings of the conference on fairness, accountability, and transparency},
  pages={329--338},
  year={2019}
}

@article{open2015estimating,
  title={Estimating the reproducibility of psychological science},
  author={Open Science Collaboration},
  journal={Science},
  volume={349},
  number={6251},
  pages={aac4716},
  year={2015},
  publisher={American Association for the Advancement of Science}
}

@article{us2021census,
  title={American Community Survey 5-Year Data (2017-2021)},
  author={{U.S. Census Bureau}},
  journal={United States Census Bureau},
  year={2021}
}

@article{campbell1959convergent,
  title={Convergent and discriminant validation by the multitrait-multimethod matrix.},
  author={Campbell, Donald T and Fiske, Donald W},
  journal={Psychological bulletin},
  volume={56},
  number={2},
  pages={81},
  year={1959},
  publisher={American Psychological Association}
}

@article{bls2025earnings,
  title={Usual weekly earnings of wage and salary workers},
  author={{Bureau of Labor Statistics}},
  journal={U.S. Department of Labor},
  year={2025}
}

@article{ma2015chicago,
  title={The Chicago face database: A free stimulus set of faces and norming data},
  author={Ma, Debbie S and Correll, Joshua and Wittenbrink, Bernd},
  journal={Behavior research methods},
  volume={47},
  number={4},
  pages={1122--1135},
  year={2015},
  publisher={Springer}
}

@article{korteling2018neural,
  title={A neural network framework for cognitive bias},
  author={Korteling, Johan E and Brouwer, Anne-Marie and Toet, Alexander},
  journal={Frontiers in psychology},
  volume={9},
  pages={1561},
  year={2018},
  publisher={Frontiers Media SA}
}

@book{gazzaniga1998mind,
  title={The mind's past},
  author={Gazzaniga, Michael S},
  year={1998},
  publisher={Univ of California Press}
}

@article{asch1946forming,
  title={Forming impressions of personality.},
  author={Asch, Solomon E},
  journal={The journal of abnormal and social psychology},
  volume={41},
  number={3},
  pages={258},
  year={1946},
  publisher={American Psychological Association}
}

@article{anderson1981foundations,
  title={Foundations of information integration theory},
  author={Anderson, Norman Henry},
  year={1981}
}

@incollection{crenshaw2013demarginalizing,
  title={Demarginalizing the intersection of race and sex: A black feminist critique of antidiscrimination doctrine, feminist theory and antiracist politics},
  author={Crenshaw, Kimberl{\'e}},
  booktitle={Feminist legal theories},
  pages={23--51},
  year={2013},
  publisher={Routledge}
}

@article{willis2006first,
  title={First impressions: Making up your mind after a 100-ms exposure to a face},
  author={Willis, Janine and Todorov, Alexander},
  journal={Psychological science},
  volume={17},
  number={7},
  pages={592--598},
  year={2006},
  publisher={SAGE Publications Sage CA: Los Angeles, CA}
}

@book{brunswik2023perception,
  title={Perception and the representative design of psychological experiments},
  author={Brunswik, Egon},
  year={2023},
  publisher={Univ of California Press}
}

@article{araujo2007ecological,
  title={Ecological validity, representative design, and correspondence between experimental task constraints and behavioral setting: Comment},
  author={Araujo, Duarte and Davids, Keith and Passos, Pedro},
  journal={Ecological psychology},
  volume={19},
  number={1},
  pages={69--78},
  year={2007},
  publisher={Taylor \& Francis}
}

@article{todorov2005inferences,
  title={Inferences of competence from faces predict election outcomes},
  author={Todorov, Alexander and Mandisodza, Anesu N and Goren, Amir and Hall, Crystal C},
  journal={Science},
  volume={308},
  number={5728},
  pages={1623--1626},
  year={2005},
  publisher={American Association for the Advancement of Science}
}

@incollection{FREEMAN2020237,
    title = {Chapter Five - Dynamic interactive theory as a domain-general account of social perception},
    editor = {Bertram Gawronski},
    booktitle = {Advances in Experimental Social Psychology},
    publisher = {Academic Press},
    volume = {61},
    pages = {237-287},
    year = {2020},
    issn = {0065-2601},
    doi = {https://doi.org/10.1016/bs.aesp.2019.09.005},
    url = {https://www.sciencedirect.com/science/article/pii/S0065260119300346},
    author = {Jonathan B. Freeman and Ryan M. Stolier and Jeffrey A. Brooks},
}

@article{fisher1993social,
    author = {Fisher, Robert J.},
    title = {Social Desirability Bias and the Validity of Indirect Questioning},
    journal = {Journal of Consumer Research},
    volume = {20},
    number = {2},
    pages = {303-315},
    year = {1993},
    month = {09},
    issn = {0093-5301},
    doi = {10.1086/209351},
    url = {https://doi.org/10.1086/209351},
    eprint = {https://academic.oup.com/jcr/article-pdf/20/2/303/5074014/20-2-303.pdf},
}

@article{goffin2011relative,
    author = {Richard D. Goffin and James M. Olson},
    title ={Is It All Relative?: Comparative Judgments and the Possible Improvement of Self-Ratings and Ratings of Others},
    journal = {Perspectives on Psychological Science},
    volume = {6},
    number = {1},
    pages = {48-60},
    year = {2011},
}

@article{sanbonmatsu1990role,
  title={The role of attitudes in memory-based decision making.},
  author={Sanbonmatsu, David M and Fazio, Russell H},
  journal={Journal of Personality and social Psychology},
  volume={59},
  number={4},
  pages={614},
  year={1990},
  publisher={American Psychological Association}
}

@article{kahneman2011thinking,
  title={Thinking, fast and slow},
  author={Kahneman, Daniel},
  journal={Farrar, Straus and Giroux},
  year={2011}
}

@article{devine1989stereotypes,
  title={Stereotypes and prejudice: Their automatic and controlled components.},
  author={Devine, Patricia G},
  journal={Journal of personality and social psychology},
  volume={56},
  number={1},
  pages={5},
  year={1989},
  publisher={American Psychological Association}
}

@article{bar2007proactive,
  title={The proactive brain: using analogies and associations to generate predictions},
  author={Bar, Moshe},
  journal={Trends in cognitive sciences},
  volume={11},
  number={7},
  pages={280--289},
  year={2007},
  publisher={Elsevier}
}

@article{bls2024nurses,
  title={Employed persons by detailed occupation, sex, race, and Hispanic or Latino ethnicity},
  author={{Bureau of Labor Statistics}},
  journal={Labor Force Statistics from the Current Population Survey},
  year={2024},
  note={Table 11},
  url={https://www.bls.gov/cps/cpsaat11.htm}
}

@article{jost2004decade,
  title={A decade of system justification theory: Accumulated evidence of conscious and unconscious bolstering of the status quo},
  author={Jost, John T and Banaji, Mahzarin R and Nosek, Brian A},
  journal={Political psychology},
  volume={25},
  number={6},
  pages={881--919},
  year={2004},
  publisher={Wiley Online Library}
}

@article{palan2018prolific,
  title={Prolific. ac—A subject pool for online experiments},
  author={Palan, Stefan and Schitter, Christian},
  journal={Journal of behavioral and experimental finance},
  volume={17},
  pages={22--27},
  year={2018},
  publisher={Elsevier}
}

@article{oppenheimer2009instructional,
  title={Instructional manipulation checks: Detecting satisficing to increase statistical power},
  author={Oppenheimer, Daniel M and Meyvis, Tom and Davidenko, Nicolas},
  journal={Journal of experimental social psychology},
  volume={45},
  number={4},
  pages={867--872},
  year={2009},
  publisher={Elsevier}
}

@article{caputo2017social,
  title={Social desirability bias in self-reported well-being measures: Evidence from an online survey},
  author={Caputo, Andrea},
  journal={Universitas Psychologica},
  volume={16},
  number={2},
  pages={245--255},
  year={2017},
  publisher={Pontificia Universidad Javeriana}
}

@article{kreitchmann2019controlling,
  title={Controlling for response biases in self-report scales: Forced-choice vs. psychometric modeling of Likert items},
  author={Kreitchmann, Rodrigo Schames and Abad, Francisco J and Ponsoda, Vicente and Nieto, Maria Dolores and Morillo, Daniel},
  journal={Frontiers in psychology},
  volume={10},
  pages={2309},
  year={2019},
  publisher={Frontiers Media SA}
}

@article{kwak2019measuring,
  title={Measuring and controlling social desirability bias: Applications in information systems research},
  author={Kwak, Dong-Heon and Holtkamp, Philipp and Kim, Sung S},
  journal={Journal of the Association for Information Systems},
  volume={20},
  number={4},
  pages={5},
  year={2019}
}

@article{axt2018best,
  title={The best way to measure explicit racial attitudes is to ask about them},
  author={Axt, Jordan R},
  journal={Social Psychological and Personality Science},
  volume={9},
  number={8},
  pages={896--906},
  year={2018},
  publisher={Sage Publications Sage CA: Los Angeles, CA}
}

@article{tversky1974judgment,
  title={Judgment under Uncertainty: Heuristics and Biases: Biases in judgments reveal some heuristics of thinking under uncertainty.},
  author={Tversky, Amos and Kahneman, Daniel},
  journal={science},
  volume={185},
  number={4157},
  pages={1124--1131},
  year={1974},
  publisher={American association for the advancement of science}
}

@article{ntoutsi2020bias,
  title={Bias in data-driven artificial intelligence systems—An introductory survey},
  author={Ntoutsi, Eirini and Fafalios, Pavlos and Gadiraju, Ujwal and Iosifidis, Vasileios and Nejdl, Wolfgang and Vidal, Maria-Esther and Ruggieri, Salvatore and Turini, Franco and Papadopoulos, Symeon and Krasanakis, Emmanouil and others},
  journal={Wiley Interdisciplinary Reviews: Data Mining and Knowledge Discovery},
  volume={10},
  number={3},
  pages={e1356},
  year={2020},
  publisher={Wiley Online Library}
}

@inproceedings{zhang2018mitigating,
  title={Mitigating unwanted biases with adversarial learning},
  author={Zhang, Brian Hu and Lemoine, Blake and Mitchell, Margaret},
  booktitle={Proceedings of the 2018 AAAI/ACM Conference on AI, Ethics, and Society},
  pages={335--340},
  year={2018}
}

@article{vangiffen2022overcoming,
  title={Overcoming the pitfalls and perils of algorithms: A classification of machine learning biases and mitigation methods},
  author={Van Giffen, Benjamin and Herhausen, Dennis and Fahse, Tobias},
  journal={Journal of Business Research},
  volume={144},
  pages={93--106},
  year={2022}
}

@article{tjoa2020survey,
  title={A survey on explainable artificial intelligence (xai): Toward medical xai},
  author={Tjoa, Erico and Guan, Cuntai},
  journal={IEEE transactions on neural networks and learning systems},
  volume={32},
  number={11},
  pages={4793--4813},
  year={2020},
  publisher={IEEE}
}

@article{linardatos2020explainable,
  title={Explainable ai: A review of machine learning interpretability methods},
  author={Linardatos, Pantelis and Papastefanopoulos, Vasilis and Kotsiantis, Sotiris},
  journal={Entropy},
  volume={23},
  number={1},
  pages={18},
  year={2020},
  publisher={MDPI}
}

@article{li2022interpretable,
  title={Interpretable deep learning: Interpretation, interpretability, trustworthiness, and beyond},
  author={Li, Xuhong and Xiong, Haoyi and Li, Xingjian and Wu, Xuanyu and Zhang, Xiao and Liu, Ji and Bian, Jiang and Dou, Dejing},
  journal={Knowledge and Information Systems},
  volume={64},
  number={12},
  pages={3197--3234},
  year={2022},
  publisher={Springer}
}

@article{arowosegbe2023data,
  title={Data bias, intelligent systems and criminal justice outcomes},
  author={Arowosegbe, Jacob O},
  journal={International Journal of Law and Information Technology},
  volume={31},
  number={1},
  pages={22--45},
  year={2023},
  publisher={Oxford University Press UK}
}

@article{howard2018ugly,
  title={The ugly truth about ourselves and our robot creations: The problem of bias and social inequity},
  author={Howard, Ayanna and Borenstein, Jason},
  journal={Science and engineering ethics},
  volume={24},
  number={5},
  pages={1521--1536},
  year={2018},
  publisher={Springer}
}

@article{tolan2019fair,
  title={Fair and unbiased algorithmic decision making: Current state and future challenges},
  author={Tolan, Song{\"u}l},
  journal={arXiv preprint arXiv:1901.04730},
  year={2019}
}

@unpublished{chen2023review,
  title={A Review Examining Biases in Workplace Hiring and Promotion Processes},
  author={Chen, Claire},
  note={The Claremont Colleges},
  year={2023}
}

\section*{Acknowledgments}
We acknowledge research funding from the Digital Communication Methods Lab at the Amsterdam School of Communication Research and the Social \& Behavioural Data Science Centre at the University of Amsterdam.

\end{document}